\documentclass[conference]{IEEEtran}
\usepackage{times}

\usepackage{amsmath}
\usepackage{amsthm}
\usepackage{amssymb}
\usepackage[utf8]{inputenc}
\usepackage[T1]{fontenc}
\usepackage{graphicx}
\usepackage{enumerate}
\usepackage{xcolor}
\usepackage{comment}
\usepackage[numbers]{natbib}
\usepackage{multicol}
\usepackage{hyperref}
\usepackage{tabularx}
\usepackage{makecell}
\usepackage{colortbl}
\usepackage{tablefootnote}
\usepackage{multirow}
\usepackage{paralist}
\usepackage{placeins}
\usepackage{cleveref}
\usepackage{nameref}

\usepackage[caption=false]{subfig}
\renewcommand{\figurename}{Fig.}
\usepackage[ruled,vlined]{algorithm2e}

\usepackage{cleveref}

\usepackage[nolist]{acronym}

\usepackage{subfig}

\usepackage{xcolor}

\usepackage{amsthm}

\newtheorem{lemma}{Lemma}
\newtheorem{theorem}{Theorem}[section]
\newtheorem{definition}{Definition}[section]
\newtheorem{proposition}[theorem]{Proposition}
\newcommand{\vx}[0]{\mathbf{x}}
\newcommand{\vu}[0]{\mathbf{u}}
\newcommand{\vw}[0]{\mathbf{w}}

\newcommand{\vv}[0]{\mathbf{v}}

\newcommand{\vxdot}[0]{\dot{\vx}}
\newcommand{\ve}[0]{\mathbf{e}}

\newcommand{\argmin}{\operatornamewithlimits{argmin}}

\newcommand{\rT}[0]{{\textrm{T}}}

\newcommand{\xtraj}[0]{\mathbf{x}_{\tau}}
\newcommand{\utraj}[0]{\mathbf{u}_{\tau}}

\newcommand{\real}[0]{\mathbb{R}}
\newcommand{\domain}[0]{\mathcal{D}}
\newcommand{\safeset}[0]{\mathcal{C}}

\newcommand{\Pb}[0]{\mathbb{P}}
\newcommand{\Qb}[0]{\mathbb{Q}}

\newcommand{\ExP}[2]{\mathbb{E}_{#1}\big[ #2 \big]}

\newcommand{\pluseq}{\mathrel{+}=}

\begin{acronym}
\acro{MPPI}{Model-Predictive Path Integral Control}
\acro{Tube-MPPI}{Tube Model-Predictive Path Integral Control}
\acro{RMPPI}{Robust Model-Predictive Path Integral Control}
\acro{CCM}{Control Contraction Metrics}
\acro{MPC}{Model-Predictive Control}
\acro{IS}{Importance Sampling}
\acro{KL}{Kullback–Leibler}
\acro{GT-ARF}{Georgia Tech Autonomous Racing Facility}
\acro{iLQG}{iterative Linear Quadratic Gaussian}
\acro{AIS}{Augmented Importance Sampling}
\acro{SAIS}{Safe Augmented Importance Sampling}
\acro{CBFs}{Control Barrier Functions}
\acro{SA-RMPPI}{Safety Augmented Robust Model Predictive Path Integral Control}
\end{acronym}

\hypersetup{
    colorlinks=true,
    linkcolor=blue,
    filecolor=magenta,      
    urlcolor=cyan,
}

\begin{document}

\title{Safety in Augmented Importance Sampling: Performance Bounds for Robust-MPPI Using Barrier States}

% You will get a Paper-ID when submitting a pdf file to the conference system
%\author{Author Names Omitted for Anonymous Review. Paper-ID [237]}

\author{\authorblockN{Manan Gandhi\IEEEauthorrefmark{1}, Hassan Almubarak\IEEEauthorrefmark{1}, Yuichiro Aoyama\IEEEauthorrefmark{1}, Evangelos Theodorou\IEEEauthorrefmark{1}}
\authorblockA{
\IEEEauthorrefmark{1}Georgia Institute of Technology\\
mgandhi@gatech.edu}}

% avoiding spaces at the end of the author lines is not a problem with
% conference papers because we don't use \thanks or \IEEEmembership

%Homer Simpson\authorrefmark{2},
%Montgomery Scott\authorrefmark{3} and
%Eldon Tyrell\authorrefmark{4}}
%Telephone: (800) 555--1212, Fax: (888) 555--1212}
%\authorblockA{\authorrefmark{4}Tyrell Inc., 123 Replicant Street, Los Angeles, California 90210--4321}}

\maketitle

\begin{abstract}
This work explores the nature of augmented importance sampling in safety-constrained model predictive control problems. When operating in a constrained environment, sampling based model predictive control and motion planning typically utilizes penalty functions or expensive optimization based control barrier algorithms to maintain feasibility of forward sampling. In contrast the presented algorithm utilizes discrete embedded barrier states in augmented importance sampling to apply feedback with respect to a nominal state when sampling. We will demonstrate that this approach of safety of discrete embedded barrier states in augmented importance sampling is more sample efficient by metric of collision free trajectories, is computationally feasible to perform per sample, and results in better safety performance on a cluttered navigation task with extreme un-modeled disturbances. In addition, we will utilize the theoretical properties of augmented importance sampling and safety control to derive a new bound on the free energy of the system.
\end{abstract}

\IEEEpeerreviewmaketitle

\section{Introduction}
Safety-critical control is a fundamental problem in dynamical systems, with many problems in robotics, healthcare, and aviation requiring safe operation. In the field of terrestrial and aerial agility there is space to balance the effort between a controller which is safe and risk averse, versus a controller that can execute aggressive maneuvers and recover gracefully while maintaining safety. The latter controller is not risk seeking, but has a built-in recovery mechanism. In this work we will present \ac{SAIS} to achieve this goal. \ac{SAIS} is connected to model-based safe-control, safe sampling, and safe reinforcement learning. 

In the context of safe control, we review some candidate techniques for maintaining the safety of a known system model through feedback. This feedback forms the lynch pin of \ac{SAIS}, as the feedback mechanism in question will be applied on each sample individually. Potential field methods utilize attractive forces for a given goal state and repulsive forces generated by obstacles. The difficulty in potential field methods lies in optimizing the ratio between the two forces, additionally, the method itself is subject to becoming trapped before narrow passages \cite{koren1991potential}.  \citet{singletary2020comparative} demonstrates that potential functions are a subset of Control Barrier Functions, whereas Control Barrier Functions can be utilized in a more general fashion as a safety filter. \ac{CBFs} have been widely successful in tasks such as bipedal walking \cite{agrawal2017discrete, ames2019control} and automatic lane keeping \cite{ames2016control}, however for complex constraints, the control is the result of a optimization scheme that is computationally expensive to run in real-time on each sample. Additionally, a particular structure is required to handle systems with high relative degree \cite{xiao2019control} and in the context of discrete \ac{CBFs}, even linear systems with linear constraints can result in a non-convex optimization problem \cite{agrawal2017discrete}. Even in the realm of sampling-based control, \ac{CBFs} can offer a useful method to improve sample efficiency, but is held back again due to computational complexity \cite{tao2021control}. Embedded barrier state (BaS) methods developed by \citet{almubarak2022safety} append the state space with the state of a barrier function, then utilize a Lyapunov stability criterion, satisfied by an optimal controller, to ensure safety through stabilization of the appended model. Through similar arguments, \citet{almubarak2022safeddp} proposes using discrete time barrier states in trajectory optimization settings and shows that the safety embedding technique exhibits great performance improvements over penalty methods which, implicitly, result in similar cost functions. This method is applicable to general nonlinear systems, and can combine a variety of nonlinear constraints at the expense of sensitivity to gradients of the barrier state dynamics, time discretization, and model error. In the proposed work, we utilize discrete embedded barrier state control as a means of safety in sampling, due to relaxation in problem assumptions, computational complexity, and ability to combine a large number of disjoint unsafe regions.

Safety in sampling based model predictive control has received recent attention \cite{martin2019trustregions, zeng2021safety}, due to the high performance capabilities of sampling-based MPC \cite{williams2017information}. Additionally, sampling based trajectory optimization shares a close connection with safe model-based reinforcement learning, which can be loosely categorized into two categories: safety within the optimization metric, and safety in sample exploration \cite{garcia2015comprehensive}. Safety in importance sampling falls into the latter category, in the same vein as the work by \citet{berkenkamp2017safe} which utilizes Lyapunov-based stability metrics to explore the policy space, and \citet{thananjeyan2021abc}, who utilize sampling to iteratively improve the policy for a nonlinear system. In the attempt to unify the safety critical control and sampling-based MPC, current attempts fall short in terms of computational efficiency, sample efficiency, and theoretical guarantees. In this paper we present a novel method of incorporating safety into augmented importance sampling. Utilizing the framework presented in \citet{gandhi2021robust}, we are able to incorporate a variety of safety-critical control techniques to enable sample efficiency, and improve the safety-task violation rate of each sample even with large unmodeled disturbances to the system. We show \textbf{three key contributions} in this work:
\begin{enumerate}
    \item Demonstration of discrete embedded barrier states as a superior method for safety-critical control in augmented importance sampling.
    \item Derivation of a new bound for \ac{RMPPI} performance that can be localized and improved through safety-guarantees.
    \item Application of safety-critical control in sampling for navigating through a cluttered environment.
\end{enumerate}

\section{Mathematical Background}

%  \epsilon_k, $\epsilon \sim \gaussian{0}{\Sigma}$, $\epsilon \in \real^m$, $\Sigma \in \real^{m \times m}$

\subsection{Problem Formulation}
Consider the discrete, nonlinear dynamical system in \Cref{eq: dynamics}:
\begin{align}
    \vx_{k+1} &= F(k, \vx_k, \vu_k ) + \vw_k, \label{eq: dynamics}
\end{align}
where, at a time step $k \in \real^+$, $\vx \in \domain \subset \real^n$, $\vu \in \real^m$, $\vw \in \real^n$, $|\vw| < D$, $D > 0$, and $F: (\real^+ \times \real^n \times \real^m) \rightarrow \real^n$. Within the domain $\domain$, we can define a \textit{safe set} $\safeset \subset \domain$. The goal of the model predictive controller will be to achieve an objective in finite time while keeping the state trajectory, $\xtraj=\{ \vx_0, \vx_1, ..., \vx_T \}$ inside the safe set $\safeset$. Specifically, the model predictive control problem considers minimizing the cost functional % $J$ in \Cref{eq: generic cost}:
\begin{align}
    J(\xtraj, \utraj) &= \phi(\vx_T) + \sum_{k=0}^{T-1} \big( q(\vx_k, k) + \lambda \vu^{\rT} \Sigma^{-1} \vu \big), \label{eq: generic cost}
\end{align}
subject to safety state constraints and control limits with $\utraj=\{ \vu_0, \vu_1, ..., \vu_{T-1} \}$. Here $\phi$ is a terminal cost, and $q$ is a nonlinear, potentially time varying state cost with a quadratic control cost penalty. $\lambda$ is known as the inverse temperature and $\Sigma \in \real^{m \times m}$ is a positive definite control penalty matrix. For the safety state constraints, we consider the superlevel set $\safeset \subset \real^n$ defined by a continuously differentiable real valued function $h: \domain\subset \mathbb{R}^n \rightarrow \mathbb{R}$ such that the set $\safeset$, its interior $\safeset^{\circ}$, and boundary $\partial \safeset$ are defined respectively as
\begin{align}  \label{eq: control barrier safe set} 
\begin{split}
    \safeset &= \{\vx  \in \domain: h(\vx) \geq 0\},\\
    \safeset^{\circ} &= \{\vx  \in \domain: h(\vx) > 0\},\\
    \partial \safeset &= \{\vx  \in \domain: h(\vx) = 0\}.
\end{split}
\end{align}
To ensure safety, the safe set $\safeset$ needs to be rendered controlled forward invariant. The notion of \textit{forward invariance} of the safe set $\safeset$ with respect to the safety-critical dynamical system \eqref{eq: dynamics} can be formally defined as follows:
\begin{definition} The set $\safeset \subset \real^n$ is said to be \textit{controlled forward invariant} for the dynamical system $\vx_{k+1} = f(k, \vx_k, \vu_k )$, if, for all $x_0 \in \safeset$, there exists a feedback control $\vu_k = K_{\text{safe}}(k, \vx_k)$, such that $\vx_{k+1} = f(k, \vx_k, K_{\text{safe}}(k, \vx_k)) \in \safeset$, for all $k \in [t_0, T]$.
\end{definition}
Notice that the definition of controlled forward invariance assumes that that the dynamics are evolving without the unknown disturbance $\vw \in \real^n$, this assumption is important when developing a safety mechanism to use in \ac{AIS}, and the framework of \ac{RMPPI} will allow us to address the disturbances caused by $\vw$. Next, we briefly review three methods for safety critical control and their applicability in sampling-based model predictive control.

\subsection{Safety Critical Control}
\subsubsection{Control Barrier Functions}
Motivated by barrier functions in optimization literature, barrier certificates \cite{prajna2003barrier, prajna2004safety} were developed as a viable method of model validation and set invariance certification. Influenced by the development of Control Lyapunov Functions (CLFs) for stability and forward invariance, Control Barrier Functions (CBFs) were proposed by \citet{wieland2007constructive}. Classically, barrier certificates and CBFs in \cite{romdlony2014uniting,ames2014control} use barrier functions similar to those popular in optimization. The fundamental idea is to render the decay rate of some barrier function, $h$, negative definite in duality to Lyapunov arguments. This enforces forward invariance of the safe set. Later, Zeroing Control Barrier Functions (ZCBF) were proposed which aim to directly ensure positive definiteness of the safety function by making its rate of change positive definite, then relaxing the constraint to allow a \textit{limited} negative rate of change \cite{ames2016control}. We formally define the Continuous Control Barrier Function in the following definition \cite{ames2019control}.

\begin{definition} For the control system: $\vxdot =  f(\vx) + g(\vx)\vu$, let $\safeset \subset \domain \subset \real^n$ be the super level set of a continuously differentiable function $h: \domain \rightarrow \real$. $h$ is a \textit{control barrier function} if there exists an extended class $\mathcal{K}_\infty$ function $\alpha$ such that $ \sup_{\vu \in \real^m} \big[ L_fh(\vx) + L_gh(\vx)\vu \big] \geq -\alpha(h(\vx))$. The set of control values that enforce the controlled forward invariance of $\safeset$ is the set:
\begin{align} \label{eq: control continuous cbf}
    K_{\text{cbf}}(\vx) = \big\{ \vu \in \real^m :  L_fh(\vx) + L_gh(\vx)\vu +\alpha(h(\vx)) \geq 0  \big\}.
\end{align}
\end{definition}

\textit{Given} a barrier function $h$, continuous time \ac{CBFs} can often be solved in a closed-form for simple linear constraints and solved with quadratic programs (QPs) for more complex constraints. Unfortunately, one of the difficulties of control barrier functions is the search for an appropriate $h$, along with solving and ensuring the feasibility of the accompanied optimization problem. Moreover, similar to CLFs, restrictions in the form of high relative degree appear. Under mild assumptions, these restrictions can be overcome with Exponential \ac{CBFs}.  Exponential \ac{CBFs} define the dynamics of the safe set function $h(x)$ as a linear system that is controlled such that the closed loop system matrix has strictly negative eigenvalues, and the initial condition satisfies Theorem 8 in \citet{ames2019control}. This method can be combined with Control Lyapunov functions to simultaneously optimize task performance and manage safety. The algorithm comes at the cost of solving a QP at each timestep, which is computationally infeasible to perform in sampling-based control.

\subsubsection{Embedded Barrier States} 
\citet{almubarak2022safety} proposed embedded \textit{barrier states} (BaS) to transform the safety objective into a performance objective by embedding the state of the barrier into the model of the safety-critical system. The barrier function's rate of change is controlled along the system's state in lieu of enforcing this rate through an inequality hard constraint. In \cite{almubarak2022safety}, the barrier state embedded model, referred to as the safety embedded model, is asymptotically stabilized, which implies safety due to boundedness of the barrier state. Next, discrete barrier states (DBaS) were proposed to perform trajectory optimization \cite{almubarak2022safeddp}, which greatly simplified the problem formulation. Namely, for the undisturbed system in \eqref{eq: dynamics} and the safety constraint \eqref{eq: control barrier safe set}, the barrier function $B:\safeset^{\circ} \rightarrow \mathbb{R}$ is defined over $h$. A key point in the definition of the barrier function $B\big(h(\vx_k)\big)$, is that $B \rightarrow \infty $ as $\vx_k \rightarrow \partial\safeset$. With this, a barrier function over $\vx$ can be defined as $\beta(\vx_k) := B\big(h(vx_k)\big)$. In the discrete settings, the barrier state is simply constructed to be
\begin{equation} \label{eq: discrete barrier state}
    \beta_{k+1} = B \circ h \circ F(k, \vx_k, \vu_k ).
\end{equation}
For multiple constraints, multiple barrier functions can be added to form a single barrier \cite{almubarak2022safeddp} or multiple barrier states. Then, the barrier state vector $\beta \in \mathcal{B} \subset \mathbb{R}^{n_\beta}$, where $n_\beta$ is the dimensionality of the barrier state vector, is appended to the dynamical model resulting in the safety embedded system:
\begin{equation} \label{eq: safety embedded model}
    \bar{\vx}_{k+1} = \bar{F}(k, \bar{\vx}_k, \vu_k ), 
\end{equation}
where $\bar{\vx} = [x, \ \ \beta]^\text{T}$ and $\bar{F} = [F, \ \ B \circ h \circ F]^\text{T}$.

One of the benefits of a safety embedded model is the direct transmission of safety constraint information to the optimal controller. This prevents two separate algorithms from \textit{fighting} one another for control bandwidth, i.e. the controller attempting to maximize performance and a safety filter attempting to maximize safety. This comes at a cost of the user having to specify the weighting between task performance and safety. For the model predictive control problem in this work, the following proposition \cite{almubarak2022safeddp} depicts the safety guarantees provided by the embedded barrier state method.
\begin{proposition} \label{prop:safety}
Under the control sequence $\utraj$, the safe set $\safeset$ is controlled forward invariant if and only if $\beta(\vx(0)) < \infty \Rightarrow \beta_k <\infty \ \forall k \in [1, T]$.
\end{proposition}
%\textcolor{red}{COMPARISONS COMPARISONS COMPARISONS COMPARISONS PARAGRAPH????}
%Finally we have embedded barrier states, but in discrete time. These system is the least restrictive of the formulations, however, the system is sensitive to discretization as it may try to "jump" the barrier.

\subsubsection{Constrained Trajectory Optimization}
Trajectory optimization techniques have also been used in conjunction with penalty methods to enforce safety. Consider a minimization problem
\begin{equation}
    \min f_{0}(\vx), \ {\text{subject to}} \ h_{0}(\vx) = -h(\vx) \leq 0.
\end{equation}
Since safety requires $h>0$, this optimization can be viewed as optimization with safety constraints. A widely used method to solve this problem is to minimize the Powell-Hestens-Rockafellar (PHR) Augmented Lagrangian (AL) associated with penalty functions \cite{Powell1969AMF, Rockafellar1974AugmentedLM, Hestenes1969MultiplierAG}:
\begin{equation}
    \min_{\vx} L_A = f_{0}(\vx) + \frac{\mu}{2}\left \Vert h_{0}(\vx) + \frac{\lambda}{\mu}\right\Vert_{+}^{2},
\end{equation}
where $\mu > 0$ and $\lambda \geq 0$ are the penalty parameter and Lagrange multiplier, respectively, and $||t||^{2}_{+} = \max(0, t^{2})$, with $t \in \mathbb{R}$. This problem can be solved iteratively by repeating the following two steps. In the first step, $L_{A}$ is approximately minimized over $\vx$, and in the second step, $\lambda$ and $\mu$ are updated. The update law of $\lambda$ can be obtained by comparing first order optimality condition of Lagrangian, $L = f_{0} + \lambda h_{0}$, and that of $L_{A}$:
\begin{align*}
    \textstyle{\frac{\partial L}{\partial \vx}} & = \textstyle{\frac{\partial f_{0}}{\partial \vx}} + \lambda \textstyle{\frac{\partial h_{0}}{\partial \vx}},\\
    \textstyle{\frac{\partial L_{A}}{\partial \vx}} &= \textstyle{\frac{\partial f_{0}}{\partial \vx}} + (\lambda + \mu h_{0})\textstyle{\frac{\partial h_{0}}{\partial \vx}},
\end{align*}
From these two equations the update law can be deduced as
\begin{equation*}
    \lambda \leftarrow \max(0, \lambda + \mu h_{0}).
\end{equation*}
The penalty parameter $\mu$ is monotonically increased when improvement of constraint violation is not satisfactory. DDP and iLQG can be used to solve the first step of the problem of dynamical systems \cite{Plancher2017ALDDP}.

\noindent \textbf{Comparison Example (Point Mass Omnidirectional Robot):}
In this example, we consider a point mass omnidirectional robot with actuation limits $[-15,15] \ \text{m}/\text{s}$ to track a given unsafe trajectory which emulates a potentially unsafe sample from MPPI. We add zero mean independent and identically distributed (i.i.d.) Gaussian noise with standard deviation of $10$ in the control channel. This simulates the exploration noise utilized in sampling-based trajectory optimization. We aim to compare the aforementioned techniques in tracking the unsafe trajectory through a safe feedback control policy. We utilize Exponential \ac{CBFs} as a safety filter around iLQG,  barrier state embedded iLQG \cite{almubarak2022safeddp} and AL-iLQG \cite{Aoyama2021constddp} to this simple tracking problem. In the upcoming figures, the initial point and the target is shown with ``x'' and ``star'', respectively. %, then perturb the controller with noise in the control channel. 

\begin{figure}[h]
    \centering
    \subfloat{\includegraphics[trim=60 0 80 0, clip, width=0.5\linewidth]{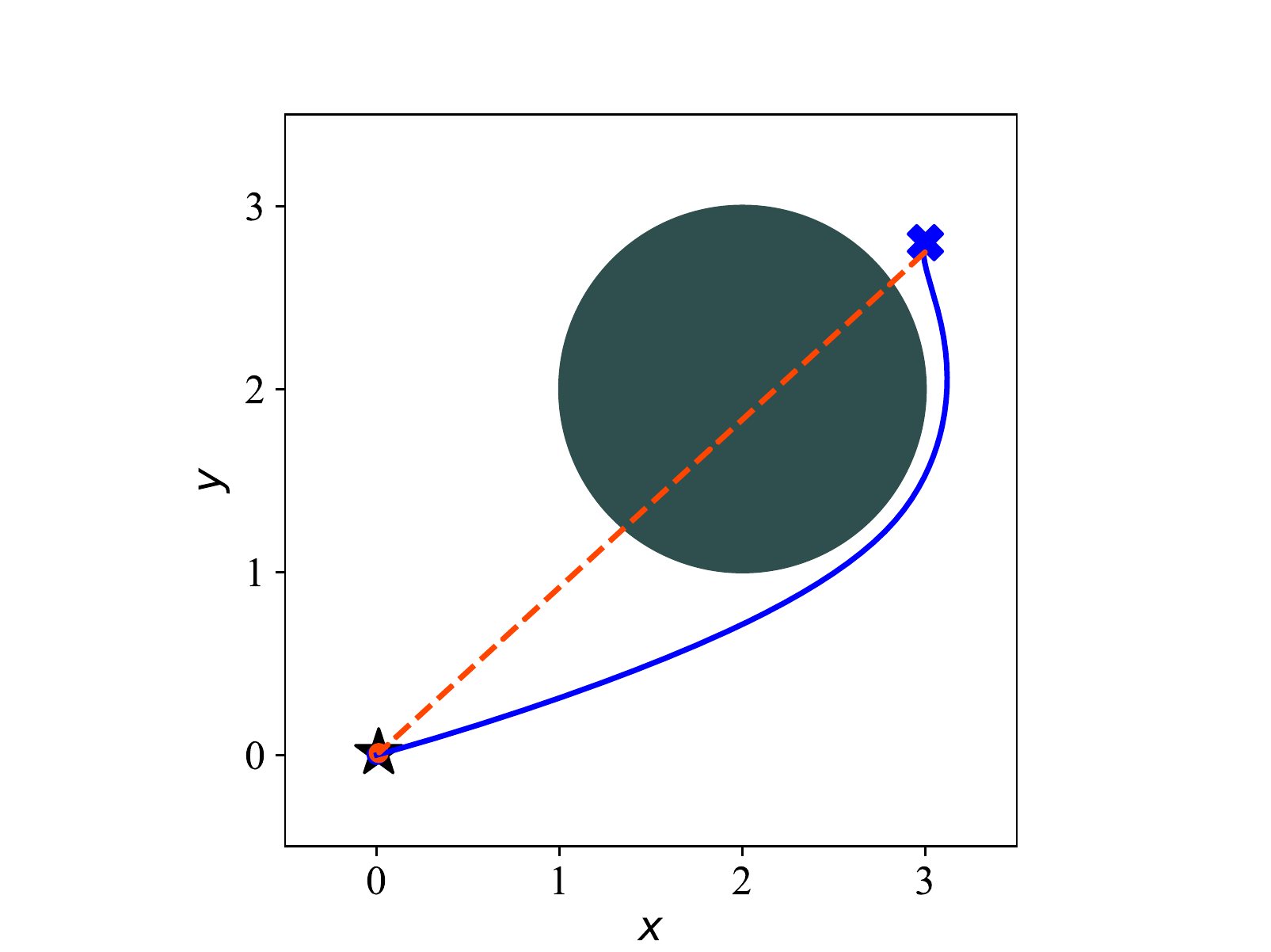}}    
    \subfloat{\includegraphics[trim=60 0 80 20, clip, width=0.5\linewidth]{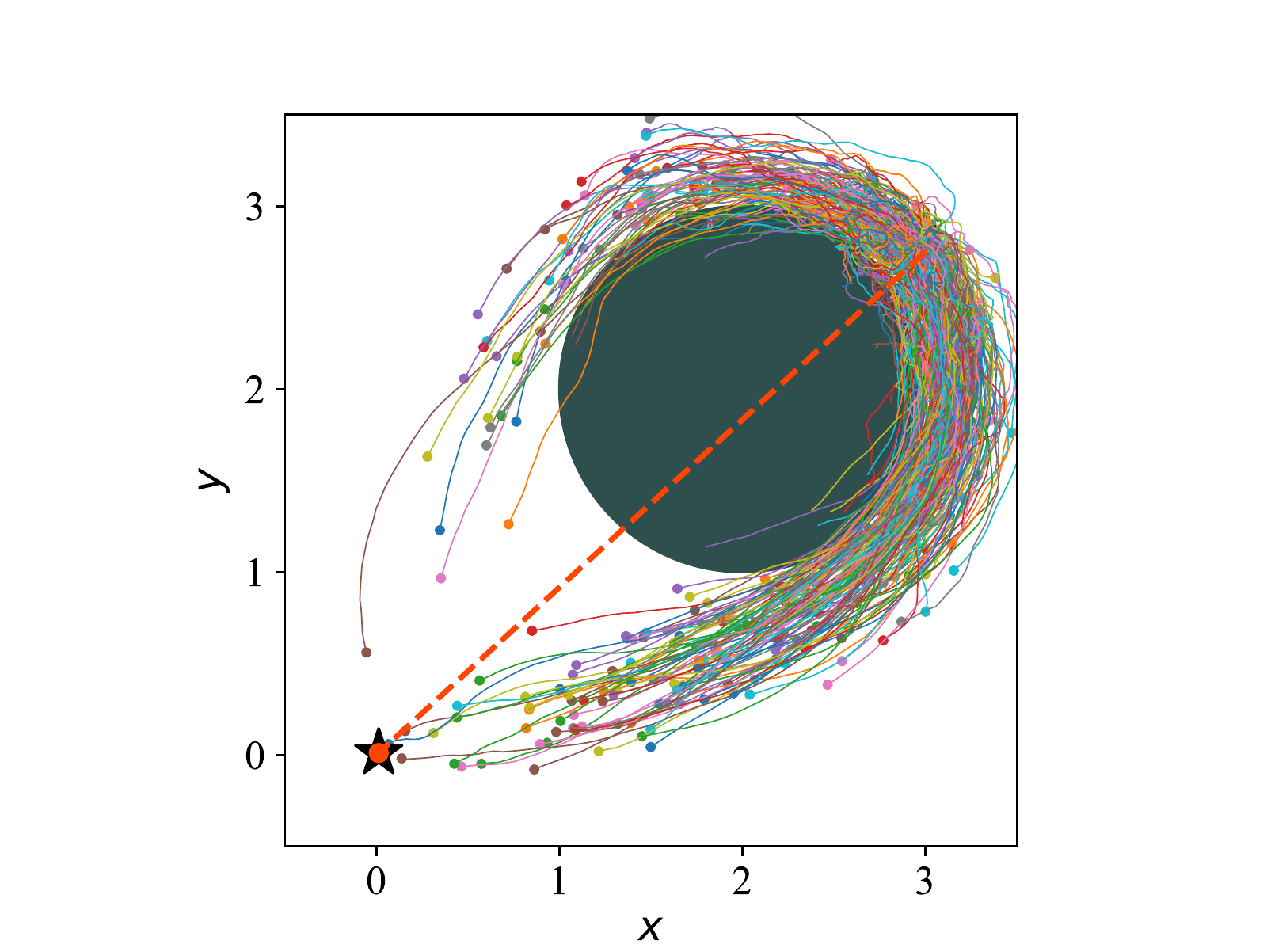}}
    \caption{Continuous Control Barrier Functions acting as a safety filter for an unsafe reference trajectory. (Left) the CBF safety filter can nominally enforce tracking of the reference while maintaining safety by solving an optimization problem at every timestep. (Right) Monte Carlo samples perturbing the nominal control make it difficult for the CBF to balance safety with target tracking, and approximately 27\% of samples are unable to remain safe and reach the target within the finite time horizon.  }
    \label{fig: CBF tracking}
\end{figure}

\begin{figure}[h]
    \centering
    \subfloat{\includegraphics[trim=60 0 80 20, clip, width=0.5\linewidth]{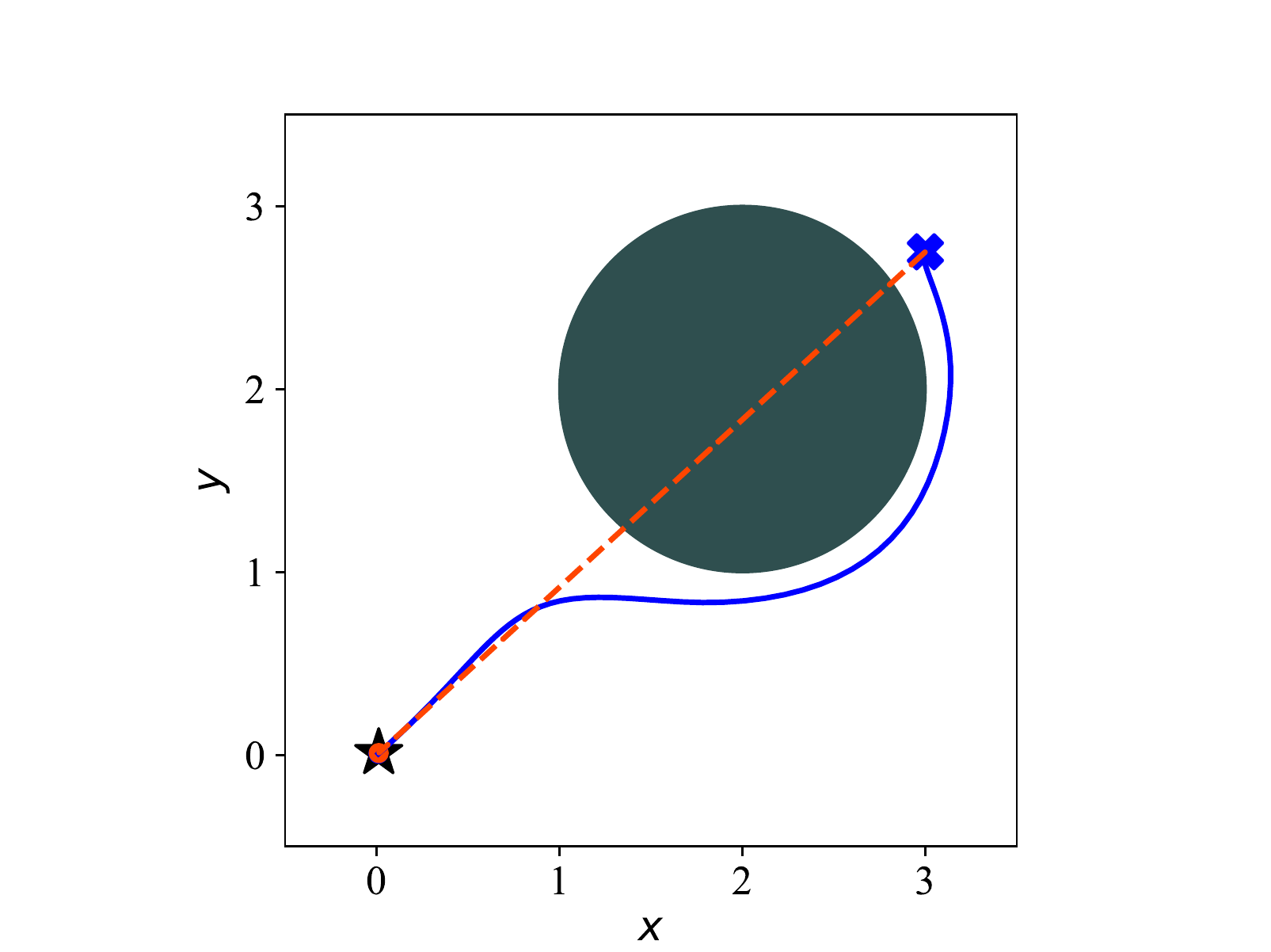}}    
    \subfloat{\includegraphics[trim=60 0 80 20, clip, width=0.5\linewidth]{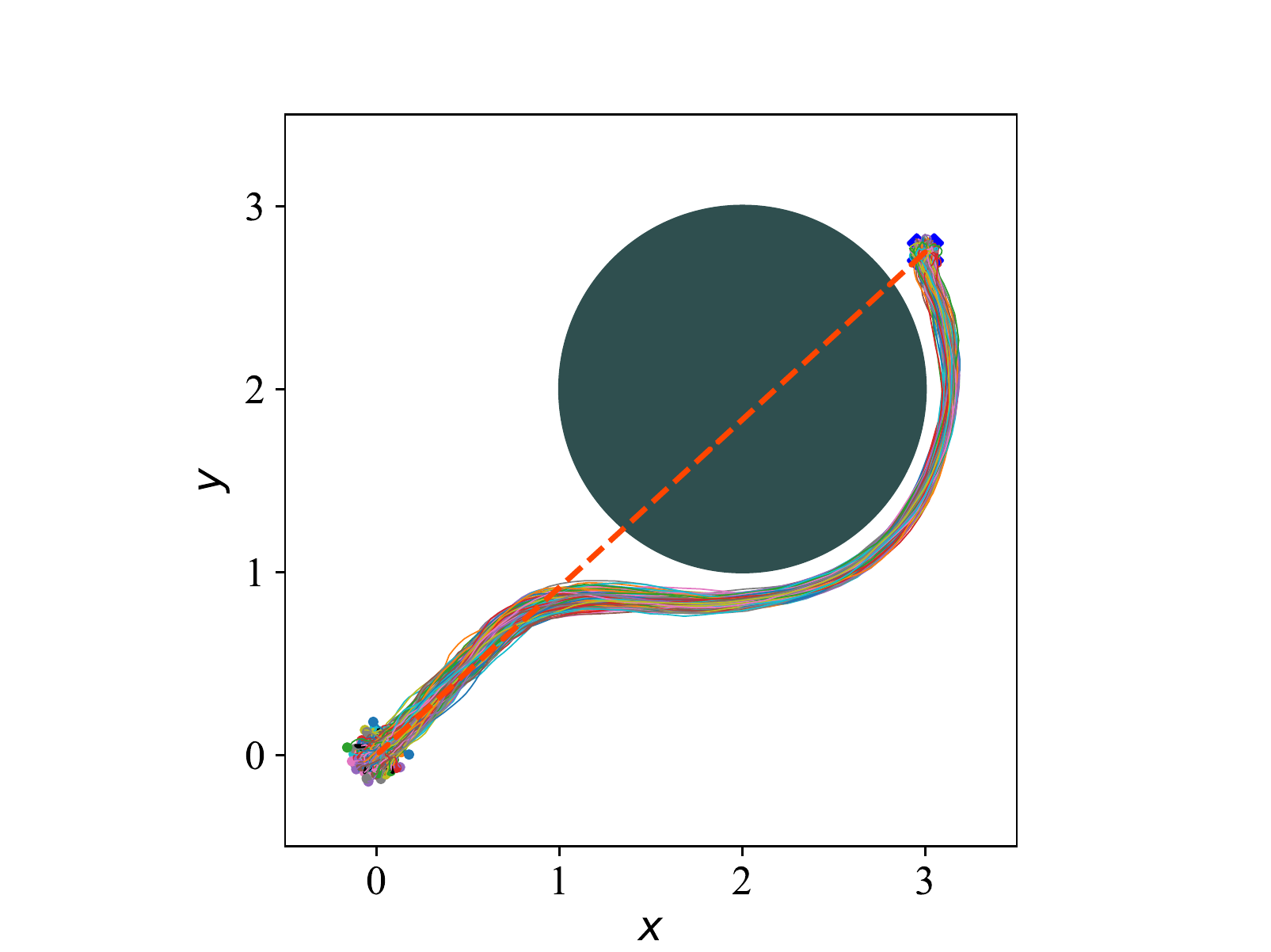}}\\
    %\subfloat{\includegraphics[trim=70 10 70 0, clip, width=0.5\linewidth]{Figures/DDP_comparisons/CBF_track_tight.eps}}    
    %\subfloat{\includegraphics[trim=70 10 70 20, clip, width=0.5\linewidth]{Figures/DDP_comparisons/CBF_noise_tight.eps}}
    \caption{BaS-iLQG tracking an unsafe trajectory (dashed orange). The generated undisturbed safe trajectory (left), shown in blue, avoids the obstacle and tracks nominal trajectory while satisfying safety. Monte Carlo simulations of $500$ samples (right) under the control noise are shown to be safe without any violation reflecting a robust and safe feedback control. %Bottom figures shows more ``safe'' trajectory obtained by penalizing safe state more. 
    }
    \label{fig: BaS-iLQG}
\end{figure}

\begin{figure}[h]
    \centering
    \subfloat{\includegraphics[trim=60 0 80 20, clip, width=0.5\linewidth]{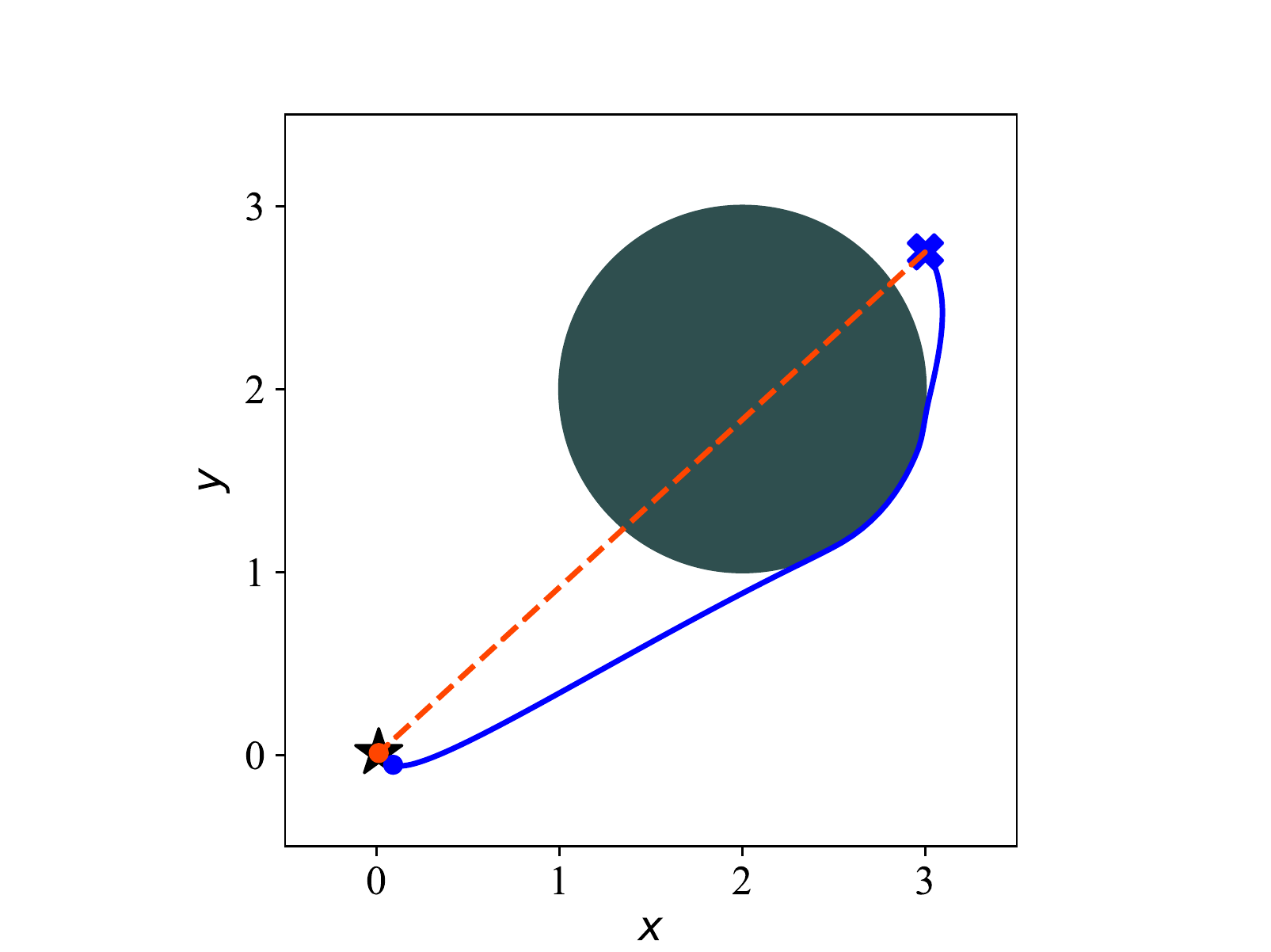}}    
    \subfloat{\includegraphics[trim=60 0 80 20, clip, width=0.5\linewidth]{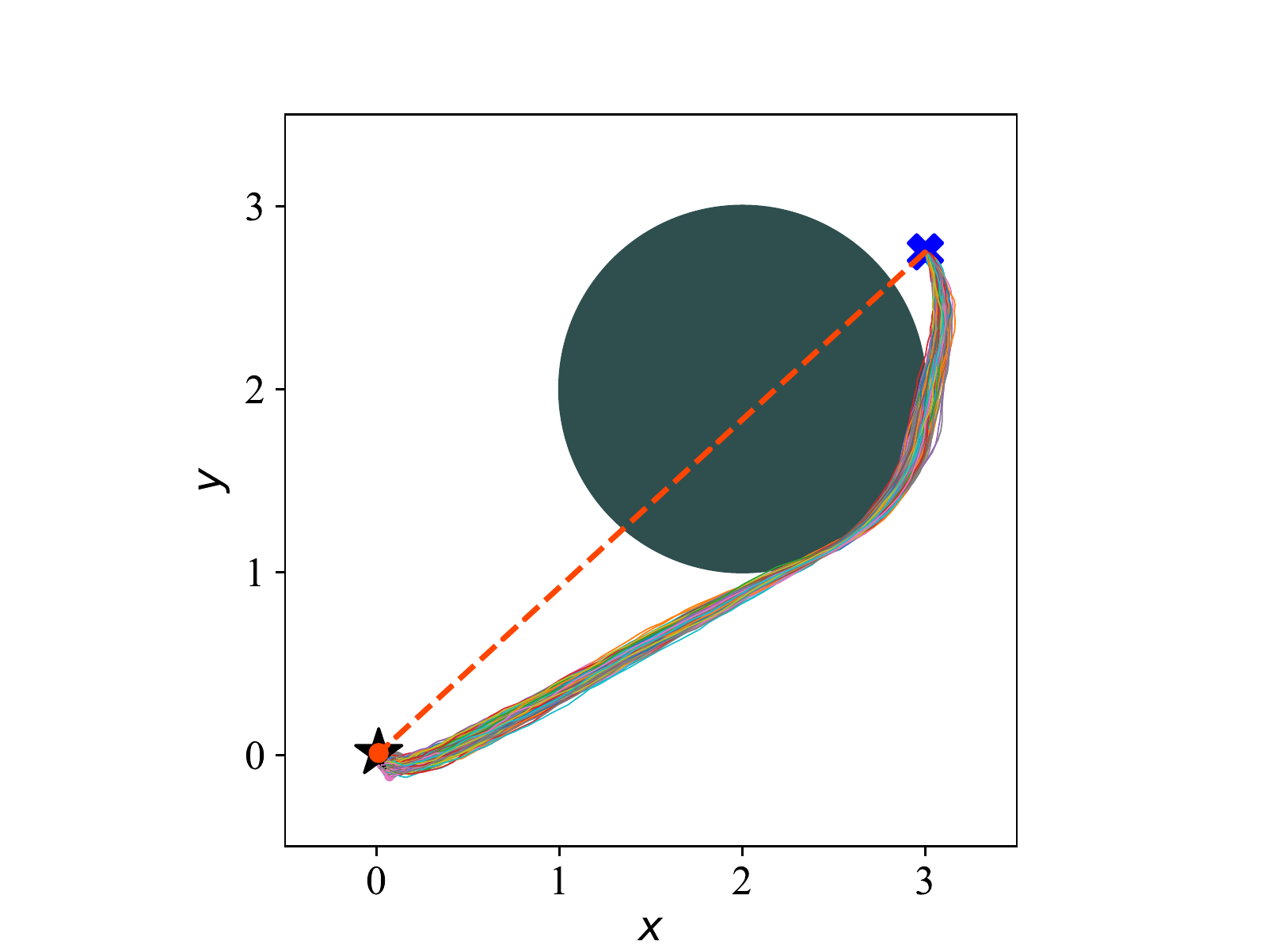}}
    \caption{A trajectory generated by AL-iLQG tracking an unsafe trajectory while dodging an obstacle (left). Trajectories obtained by Monte Carlo simulation with perturbed control (right). Due to the formulation of AL-iLQG having $h_{0}\leq 0$, the trajectory exactly meets the safety constraint. Feedback gain from iLQG ensures a tight distribution of trajectories, however approximately 60\% of trajectories are unsafe due to the proximity of the nominal trajectory to the obstacle.}
    \label{fig:AL DDP}
\end{figure}

In \Cref{fig: CBF tracking}, the ECBF safety filter is able to track to the target in the nominal case, but is unable to maintain safety with large control deviations. In \Cref{fig: BaS-iLQG}, BaS-iLQG safely handles relatively high disturbances in the control, with $0$ violations, and successfully reaches the target. Additionally, this method rapidly converges to \textit{track} the given trajectory as soon as it avoids the obstacle. Finally, in \Cref{fig:AL DDP}, the nominal trajectory is able to track the reference with minimal required deviation from the obstacle, but when perturbed with noise, this results in poor safety performance.

% COMMENT: We mention this multiple times, and we should avoid blatant favoritism in the comparisons
% This is a direct result of the barrier state impacting the value function gradient through the cost function, and the barrier state being optimized along with the system states and controls \cite{almubarak2022safeddp}.

% not only affects the cost function and can be tuned through the cost function as the other states, but affects the value function's gradient which supplies the optimal controller and the fact that the barrier state is optimized over along the system's states and controls as decision variables as discussed in . 

Discrete embedded barrier state control appears to be the least restrictive of the aforementioned algorithms in terms of both computational complexity and problem formulation, making it an ideal candidate for sampling based model predictive control and motion planning. Note that the inherent sensitivity of \textit{discrete} barrier states to time discretization may cause it to \textit{jump} the barrier. This can be partially remedied with indicator functions, i.e. crash costs, in sampling based methods to reject such violating trajectories during the optimization \cite{almubarak2022safeddp}. Utilizing crash costs to penalize unsafe trajectories is similar to penalty methods, which utilize an equivalent cost function, but as as discussed earlier, barrier states play a key role as part of the dynamics and are available directly feedback controller to perform feedback.

%%%%%%%%%%%%%%%%%%%%
\begin{comment} %% CART POLE SIMULATION

\begin{figure}[h]
    \centering
    \subfloat{\includegraphics[trim=50 10 50 0, clip, width=0.5\linewidth]{Figures/DDP_comparisons/DBaSDDP_std10_99CR.pdf}}    \subfloat{\includegraphics[trim=50 10 50 0, clip, width=0.5\linewidth]{Figures/DDP_comparisons/DBaSDDP_std10_1000samples.pdf}} \\
    \subfloat{\includegraphics[trim=50 10 50 20, clip, width=0.5\linewidth]{Figures/DDP_comparisons/DBaSDDP_std10_99CR_smallDBaScoeff.pdf}}    \subfloat{\includegraphics[trim=50 10 50 20, clip, width=0.5\linewidth]{Figures/DDP_comparisons/DBaSDDP_std10_1000samples_smallDBaScoeff.pdf}}
    \caption{DBaS-DDP tracking an unsafe trajectory (dashed red) under zero mean Gaussian noise with standard deviation of $10$ in the control inputs emulating the sampling method. The generated undisturbed safe trajectory is shown in dark while the shaded region represents $99\%$ confidence region (left). The Monte Carol simulations of $1000$ trajectories are shown in the right sub-figure. Bottom figures show less restrictive DBaS-DDP.}
    \label{fig:Cart pole}
\end{figure}
\end{comment}

%%%%%%%%%%%%%%%%%%%%%%%%

% These dynamics can be concatenated together into an augmented state $\tilde{x} = [x,z]$, which can be jointly optimized with both the trajectory objective and the safety objective. In this case $h(x)$ defines the safety set, and $B(x)$ defines the barrier function.

\section{Safety in Sampling based control}
\subsection{Information Theoretic Model Predictive Control}
We briefly review Free Energy, Relative Entropy, and the connection to model predictive control. Additional details can be found in \citet{williams2017information}. First, we define free energy with the following expression:
\begin{align}
\mathcal{F}(S, \mathbb{P}, \vx_0, \lambda) &= -\lambda \log \Bigg[ \mathbb{E}_{\mathbb{P}}(\exp(-\frac{1}{\lambda} S(V)) \Bigg], \label{eq: free energy} \\
    S(V, \vx_0) &= \phi(\vx_T) + \sum_{t = 0}^{T-1} q(\vx_t) \label{eq:pathcost}.
\end{align}
 $\lambda$ is the inverse temperature. $S$ is the cost-to-go function, which takes in initial condition $\vx_0$ and set of random variables that generate a sequence of controls $V = \{ \vv_0, \vv_1, ..., \vv_{T-1} \}$, which in turn generate a trajectory $\xtraj$ evaluated with the terminal and state cost functions $\phi$ and $q$ respectively. The probability measure $\Pb$ is utilized to sample the controls of the system when computing the free energy.

We can upper bound the free energy using Jenson's Inequality,
\begin{align}
    \text{KL}(\mathbb{Q} || \mathbb{P}) &= \mathbb{E}_{\mathbb{Q}}\Big[ \log (\frac{\text{d}\mathbb{Q}}{\text{d}\mathbb{P}}) \Big], \\
    \mathcal{F}(S, \mathbb{P}, \vx_0, \lambda) &\leq \mathbb{E}_{\mathbb{Q}}\Big[ S(V) \Big] + \lambda \text{KL}(\mathbb{Q} || \mathbb{P}). \label{eq: free energy relative entropy}
\end{align}

\Cref{eq: free energy relative entropy} now represents an optimization problem. Assume $\Qb^*$ is the optimal control distribution, and when the free energy is computed with respect to $S$ and $\Qb^*$, the free energy is minimized. This optimal free energy is upper bounded by the free energy computed from another distribution $\Qb$, summed with the KL-divergence between $\Qb^*$ and $\Qb^*$. In practice, MPPI assumes a form of the optimal distribution $\Qb^*$, which cannot be directly sampled from. Instead the KL-divergence term is utilized as an information theoretic metric that is used to drive a controlled distribution $\Qb$ closer to the optimal:

\begin{align}
    U^* = \argmin_U \text{KL}(Q^*||Q).
    \label{eq:control_obj_MPPI}
\end{align}

The authors in \citet{williams2017information} show that the solution to \Cref{eq:control_obj_MPPI} is equivalent to solving $\vu_t^* = \int q^*(V) \vv_t dV$, MPPI utilizes iterative importance sampling to approximate samples from the optimal distribution.

% The free energy bound derived in \citet{gandhi2021robust} can be used to determine how close a given system is to task failure, as well as how aggressive the tracking controller is in returning the system back to the nominal state.
\subsection{MPPI on Barrier State Embedded Dynamics}
A first step to incorporating safety into MPPI is by directly solving the model predictive control problem for the safety embedded model \eqref{eq: safety embedded model}. The constrained optimization problem is transformed to  
\begin{equation} \label{eq: cost for embedded model}
\begin{split}
    J(\bar{\xtraj}, \utraj) &= \phi(\bar{\vx}_T) + \sum_{k=0}^{T-1} \big( q(\bar{\vx}_k, k) + \lambda \vu^{\rT} \Sigma^{-1} \vu \big) \\
    \text{s.t.}   \  &\bar{\vx}_{k+1} = \bar{F}(k, \bar{\vx}_k, \vu_k ) 
\end{split} \end{equation}
Given the initial condition $\bar{\vx}_0$, initial control sequence $\utraj$, we can write the optimal control update as:
\begin{align*}
    \vu_k^* = \ExP{\Qb}{w(V) \vv_k},
\end{align*}
where $w(V) = \exp{\Big( \frac{1}{\lambda}\Big( S(V, \bar{\vx}_0) - \sum_{k=0}^{T-1} \vu^{\text{T}} \Sigma^{-1} \vv \Big) \Big)}$ are the importance sampling weights.

This formulation allows MPPI to combine the optimization of task performance with safety, however, as we will see in \Cref{section: experiments}, some limitations exist. In general, there are scenarios where we lose the ability to explore if the dynamics are too close to an obstacle or in an undesirable region of state space, demonstrated in \citet{gandhi2021robust}. Vanilla MPPI with discrete barrier state augmented dynamics is still subject to the same failure conditions as MPPI without barrier states. This motivates the next section: applying safety in the importance sampler.

% COMMENT this limitation is not explicitly shown, and instead we can discuss it as part of the experiment.
% The first is related to sample efficiency and the exponential transformation. If, during the sampling phase, of MPPI, the systems breach the barrier, the barrier state diverges and the cost of that trajectory diverges, effectively eliminating that sample by setting its importance sampling weight to zero.

\section{Safety in Augmented Importance Sampling}
In this section, we outline our primary approach to handling safety in importance sampling. We begin with the formulation of a safety embedded model, which includes an embedded barrier state, that, when bounded, ensures feasibility of the safety problem.

First we define our two systems, one real with state $\vx$, and one nominal with state $\vx^*$,
\begin{align}
    \vx^*_{k+1} &= F(k, \vx^*_k, 
    \vu^*_k +  \epsilon_k), \label{eq:nominal_dynamics} \\
    \vx_{k+1} &= F(\vx_k, \vu^*_k + \epsilon_k + K(k, \vx_k, \vx^*_k)) + \vw_k. \label{eq:real_dynamics}
\end{align}
These dynamics are identical to \Cref{eq: dynamics}, with the addition of the nominal state and tracking controller and multivariate Gaussian noise $\epsilon \in \mathcal{N}(0, \Sigma)$ acting directly on the control channel. Note that between these two systems, the same $\epsilon_t$ is applied to both systems when simulating them forward. 

The difference between the Augmented Importance Sampling in the original work, and the one described here is the underlying tracking controller. \citet{gandhi2021robust} utilized a global exponentially stable tracking controller, which is difficult to design for some nonlinear systems. In this case, we will utilize a feedback controller that assures \textit{controlled forward invariance} of the nominal trajectory within a specified safe region, i.e,  $x_t \in \mathcal{C} \subset \mathcal{D} \subset \mathcal{R}^n$, $\forall t > 0$. We will design this controller utilizing discrete embedded barrier states, with iLQG as the optimization scheme to compute feedback gains that can be applied on each sample.

Utilizing this novel scheme for enabling safety in sampling, we can provide a theorem that can be used to compute free energy growth in the system.

\begin{lemma}
Let $\mathcal{F}_{MC}(S, \Pb, \vx_0, \lambda)$ be the estimate of the free energy of the real system, $\mathcal{F}_{MC}(S, \Pb, \vx_0^*, \lambda)$ be the estimate of the free energy of the nominal system, and $E_M^V$ be the upper bound on the uncertainty of the free energy estimates. Given a nominal control sequence $\utraj^*$, define $\mathcal{B}$ to be a tube of radius $R$ defined around the nominal trajectory $\xtraj$. Define a new safeset $\safeset^* = \mathcal{B} \cup \safeset$. Let $L_q$ be the Lipshitz constant of the state cost function $q(\vx)$ in $\safeset^*$, and $L_{\phi}$ let be the Lipshitz constant of the terminal cost function $\phi(\vx)$ in $\safeset^*$. Finally, denote $\| F(\vx_0, \vu) - \vx_0\| + \|\vx_0^* - \vx_0 \| + D$ as  $D_F(\vx_0, \vx_0^*, \vu)$, and assume that there exists a safe controller $K_{\text{safe}}$ such that \Cref{eq:real_dynamics} is controlled forward invariant with respect to $\safeset^*$. Then, the growth of the free energy of the real system is bounded by:
\begin{align}
\Delta \mathcal{F}_{MC}(S, \Pb, &\vx_0, \lambda) ~\le \left(\alpha -\mathcal{F}_{MC}\left(S, \Pb, \vx_0^*, \lambda\right)\right) + 2E_M^V +\nonumber \\ &\left(R(\text{L}_{\phi} + (T-1) \text{L}_{q}\right)D_F(\vx_0, \vx_0^*, \vu) . \label{eq:bound_growth_rfe}
\end{align} 
\label{lemma:bound}
\end{lemma}
\vspace{-0.5cm}
\begin{proof}
From the a given trajectory $\xtraj$, let us analyze the cost-to-go differential between the nominal trajectory $\xtraj^*$ and $\xtraj$:
\begin{align*}
    S(V) &= \phi(\vx_T) + \sum_{k=0}^{T-1} q(\vx_k), \\
         &= \phi(\vx_T^* + (\vx_T - \vx_T^*)) + \sum_{k=0}^{T-1} q(\vx_k^* + (\vx_k - \vx_k^*))), \\
         &= \phi(\vx_T^* + \ve_T) + \sum_{k=0}^{T-1} q(\vx_k^* + \ve_k)),
\end{align*}
where $e_t$ is the difference in given state and nominal state at time $t$. Let us now take a first order Taylor expansion:
\begin{align*}
S(V) &= \phi(\vx_T^*) + \frac{\text{d} \phi}{\text{d}\vx}\Big|_{\vx^*_T}^{\text{T}} \ve_T + \sum_{k=0}^{T-1}\Big( q(\vx_k^*) + \frac{\text{d}q}{\text{d}\vx}\Big|_{\vx^*_k}^{\text{T}} \ve_k \Big) \\\nonumber &+ O(||\ve_k||^2_2).
\end{align*}
Utilizing the safe controller $K_\text{safe}$, the error dynamics can be bounded by the tube radius: $||\ve_k||_2 < R$. We can pull out the terms that represent the cost-to-go for the nominal and given trajectories, and set up an inequality by removing the higher order terms and inserting the bound on $\ve_k$ and our local Lipshitz constants $L_{\phi}$ and $L_{q}$:
\begin{align*}
    S(V, \vx_0) &\leq S(V, \vx_0^*) + (L_\phi  + (T-1) L_q) R .
\end{align*}
Analogous to the proof of lemma 3 in \citet{gandhi2021robust}, we can use the bound on the cost-to-go of the real state to bound the free energy of the real system:
\begin{align*}
\mathcal{F}(S, \Pb, \vx_0, \lambda) \le \mathcal{F}(S, \Pb, \vx_0^*, \lambda)  + \delta ,\label{eq:real_nominal_FE}
\end{align*}
where $\delta = \left(L_\phi  + (T-1) L_q \right) R$. The rest of the proof is identical to \citet{gandhi2021robust}.
\end{proof}

\Cref{lemma:bound} implies that the last term in the inequality \Cref{eq:bound_growth_rfe} can be computed using \textit{local} Lipshitz constants that are computed about each nominal trajectory. This is a key difference from the previous result, as large crash costs utilized in RMPPI forced the Lipshitz constants near obstacles to be very large. Since there was no guarantee on the domain around which each sample would explore, it was impossible to localize the bounds on a per trajectory basis, generating large upper bounds that were only meaningful near the obstacles. In contrast, this result allows us to compute local bounds at each iteration, to assess how close the real system is to task failure in other costs. Additionally, if the tube around the nominal trajectory intersects the unsafe set $\safeset$, we have additional information regarding the performance of the vehicle.

\subsection{Safety Augmented Robust Model Predictive Path Integral Control} In this section we present an algorithm for safety augmented importance sampling, which is then utilized in a framework we call \ac{SA-RMPPI}.
\begin{algorithm}[htbp!]
\footnotesize
\SetKwInOut{Input}{Given}
\Input{
      $F, K_\text{safe}, R$: Dynamics, safe controller, tube radius\;
      $q$, $\phi$, $\Sigma$, $T$, $N$: Cost function and sampling parameters\;
      $\lambda, \alpha$: Temperature and cost threshold
      }
\SetKwInOut{Input}{Input}
\Input{
    $\vx_0, \vx_0^*$, $U$, $K$: Real/nominal state, IS sequence, feedback\;
}
\BlankLine
\For{$n \leftarrow 1$ \KwTo $N$}{
    $\vx \leftarrow \vx_0$;\quad $\vx^* \leftarrow \vx_0^*$;\quad $S_n, \hat{S}_n, S^{real}_n \leftarrow 0$\;
    Sample $\mathcal{E}^n = \left( \epsilon_0^n \dots \epsilon_{T-1}^n \right), ~\epsilon_k^n \in \mathcal{N}(0, \Sigma)$\;
    \For{$k \leftarrow 0$ \KwTo $T-1$}{
        $k_{fb} \leftarrow K_\text{safe}(\vx, \vx^*)$\;
        $\vx \leftarrow F\left(\vx, \vu_k + \epsilon_k^n + k_{fb} \right)$\;
        $\vx^* \leftarrow F\left(\vx^*, \vu_k + \epsilon_k^n \right)$\;
        $\hat{S}_n \pluseq q(\vx) + \frac{\lambda (1-\beta)}{2} k_{fb}^\rT \Sigma^{-1} k_{fb}$\;
        $S_n \pluseq q(\vx^*)$ \;
        $S^{real}_n \pluseq q(\vx) + \frac{\lambda  (1-\beta)}{2}\left(\vu + k_{fb}\right)^\rT\Sigma^{-1}\left(\vu + 2\epsilon + k_{fb}\right)$
    }    
    $\hat{S}_n += \phi(\vx)$, $S_n += \phi(\vx^*)$, $S^{real}_n += \phi(\vx)$\;
    $S^{nom}_n = \frac{1}{2}S_n + \frac{1}{2}\max\left( \min \left(\hat{S}_n, \alpha \right), S_n \right)$\;
    \For{$k \leftarrow 0$ \KwTo $T-1$}{
        $S^{nom}_n \pluseq \frac{\lambda}{2}\sum_{k=0}^{T-1}\left( \vu^\rT \Sigma^{-1}\vu_k + 2\vu^\rT \Sigma^{-1}\epsilon_k^n\right) $\;
    }
}
\Return{$S^{nom}$,
        $S^\text{real}$,
        $\mathcal{E}$\;
}
\caption{Safety Augmented Importance Sampler (SAIS)} 
\label{alg: safe AIS}
\end{algorithm}

Safe Augmented Importance Sampling \Cref{alg: safe AIS}, together with the nominal state propagation and safe tracking controller, is the next step in a framework that can now optimize safety constraints while dealing with out-of-distribution disturbances from $\vw$. Note that depending on the region of attraction of the safe controller, as well as the magnitude of out-of-distribution disturbances, the theoretical bound on free energy may be violated.

\section{Experiments}
\label{section: experiments}
In this section we run a point mass omnidirectional robot with through a cluttered environment with large disturbances in the velocity term. All experiments were able to be run in real time on an Intel i5-4670K with a GTX 1080 graphics card. In these experiments, the control limitations are set to $\pm5 \ \text{m}/\text{s}$, with instantaneous velocity disturbances pulling from a Gaussian distribution with variance up to $100 \ \text{m}/\text{s}$. The aim of these comparisons is to demonstrate the capabilities of this framework to 1.) outperform MPPI simply augmenting the dynamics and optimizing the barrier state, 2.) perform safe sampling under extreme disturbance conditions that are out-of-distribution for the controller parameters for a system that is further constrained by control limits.

The experiment consists of a navigation task through an obstacle field of 30 obstacles of varying radii, creating a map that has many local minima where the system can become stuck, as well as narrow passages that must be safely traversed. The metrics in \Cref{tab: statistical comparison of point mass} are safety percentage and root mean squared error (RMSE) in meters from the goal state. The safety percentage is defined as the number of samples in the Monte Carlo study that crashed into an obstacle. In all experiments, MPPI satisfaction of the safety constraint led to the controllers \textit{reaching} the goal state within a one meter radius.

\subsection{Barrier State MPPI  (BaS-MPPI)}
\begin{figure}[h]
    %\hspace{-2.5mm}
    \subfloat{\includegraphics[trim=25 5 20 30, clip, width=.55\linewidth]{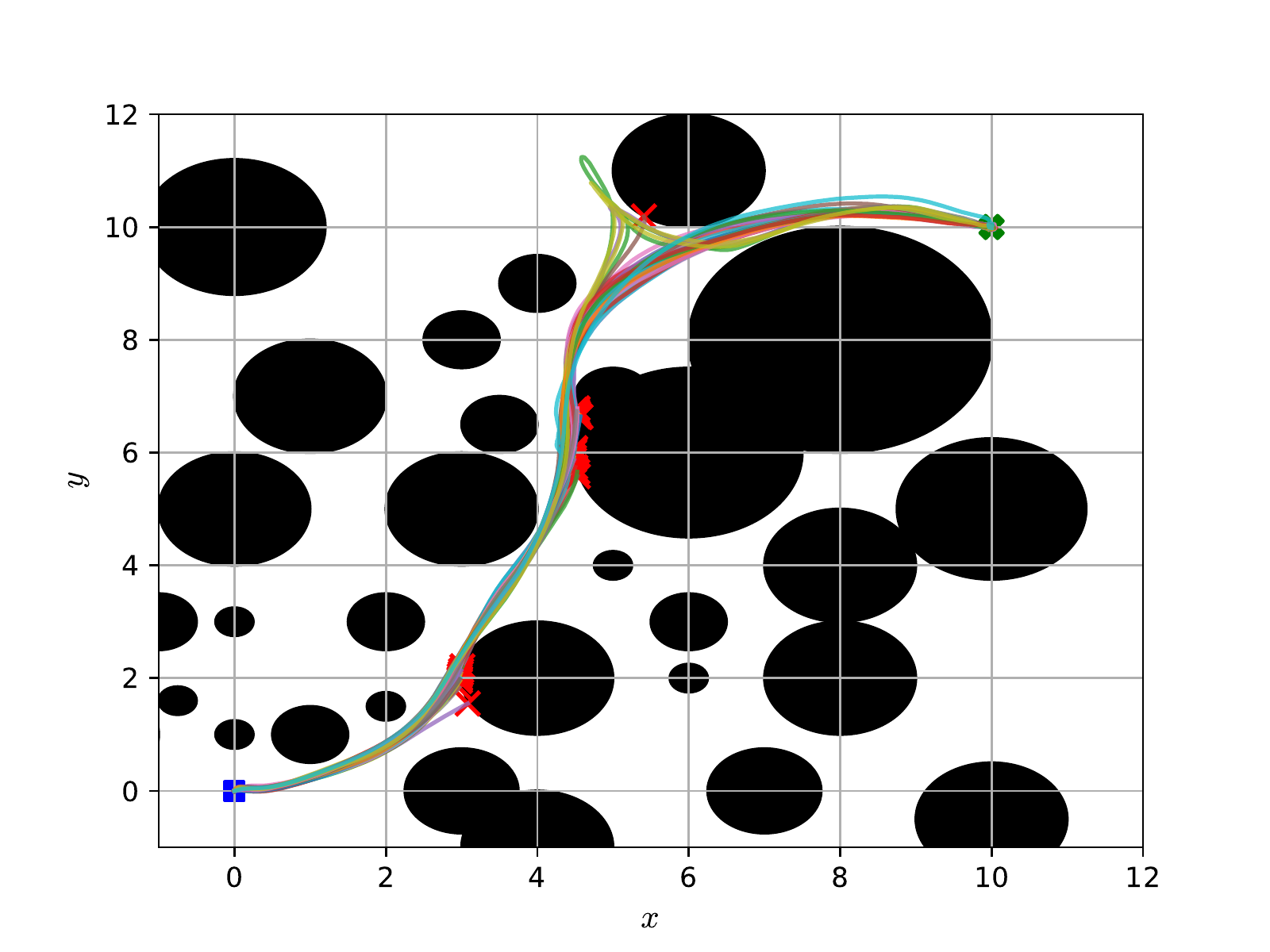}}
    %\hspace{-7mm}
    \subfloat{\includegraphics[trim=20 5 20 30, clip, width=.55\linewidth]{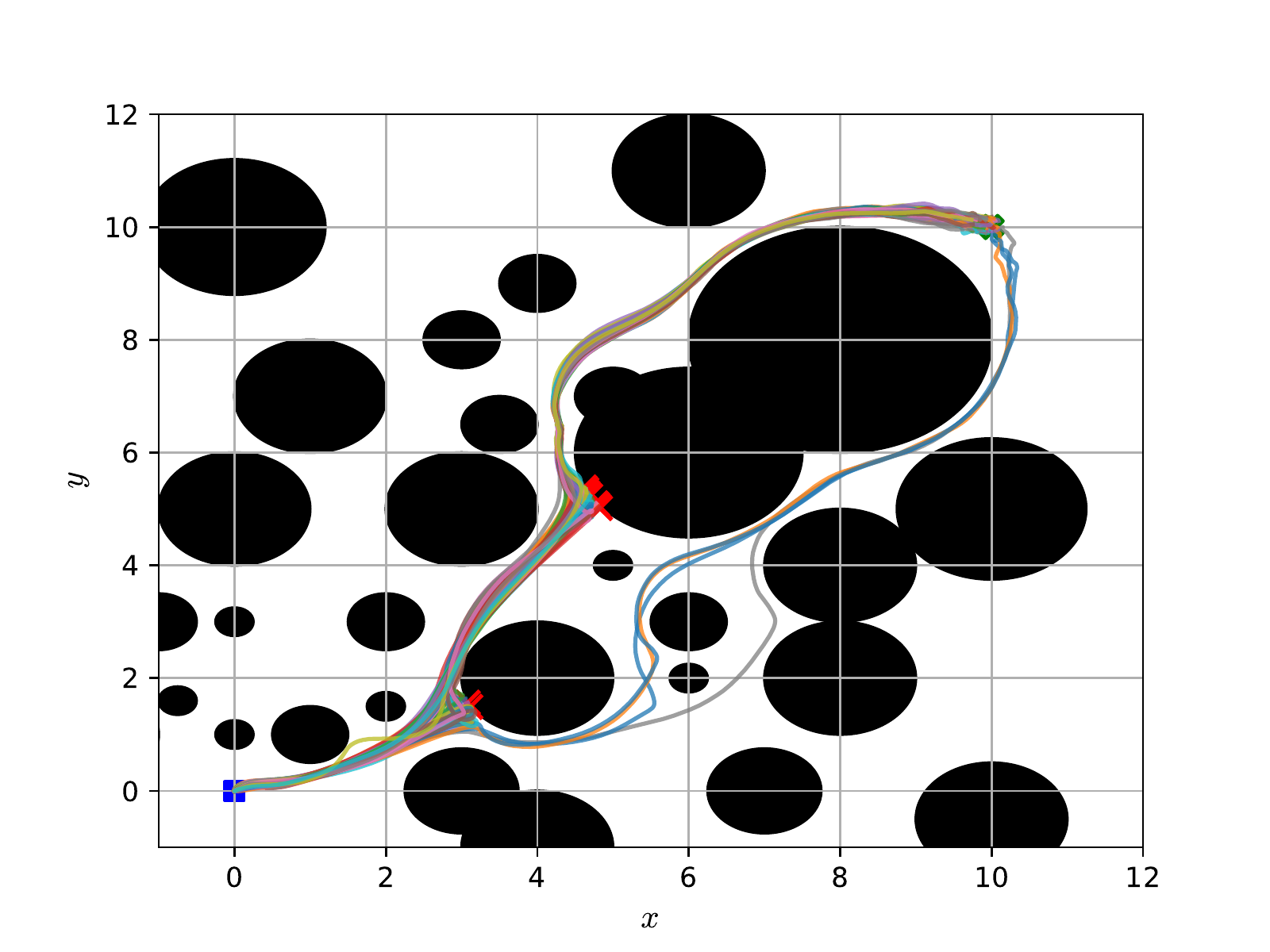}}
    \\
    %\hspace{-2.5mm}
    \subfloat{\includegraphics[trim=25 5 20 30, clip, width=.55\linewidth]{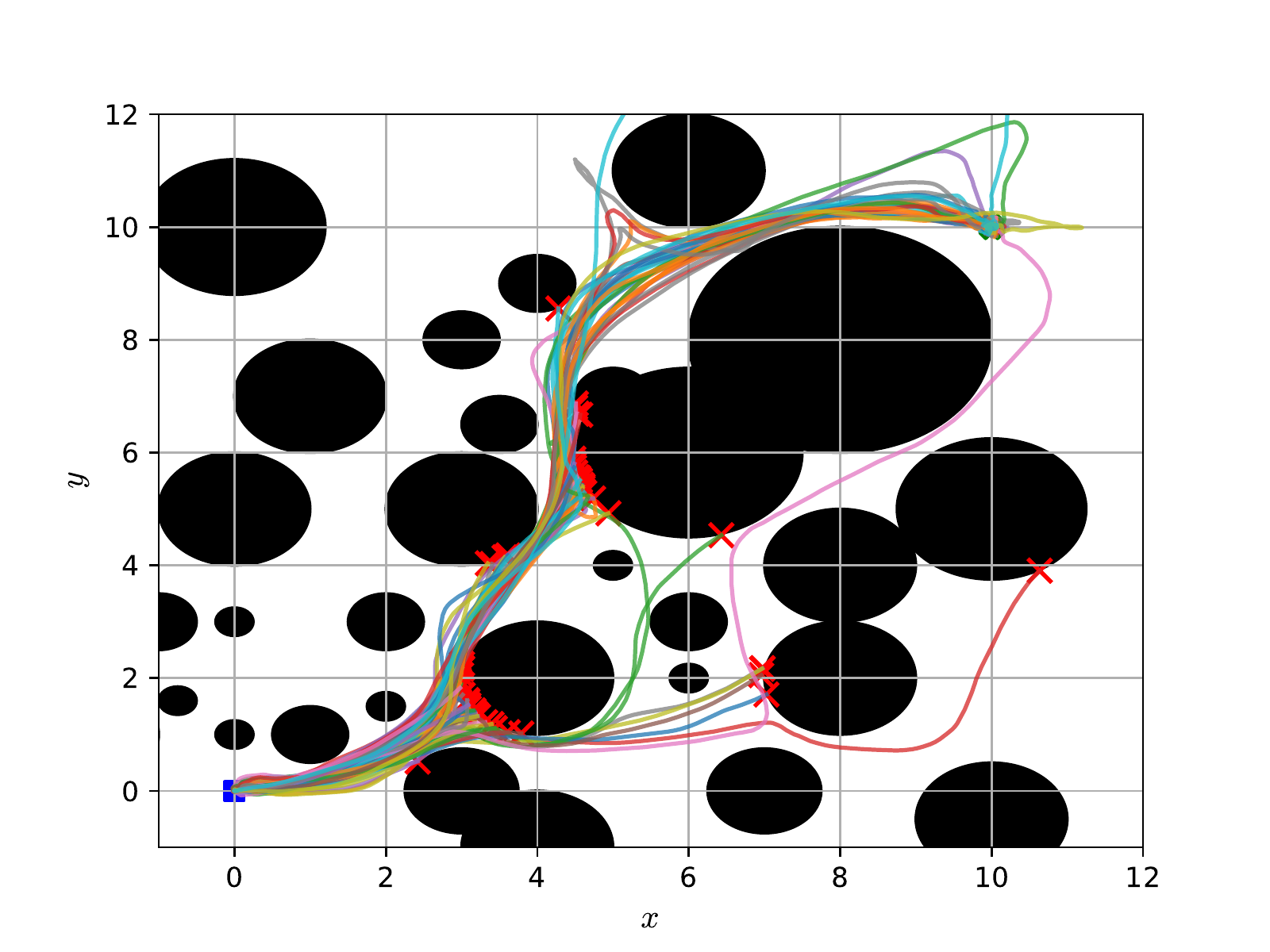}}
    %\hspace{-7mm}
    \subfloat{\includegraphics[trim=20 5 20 30, clip, width=.55\linewidth]{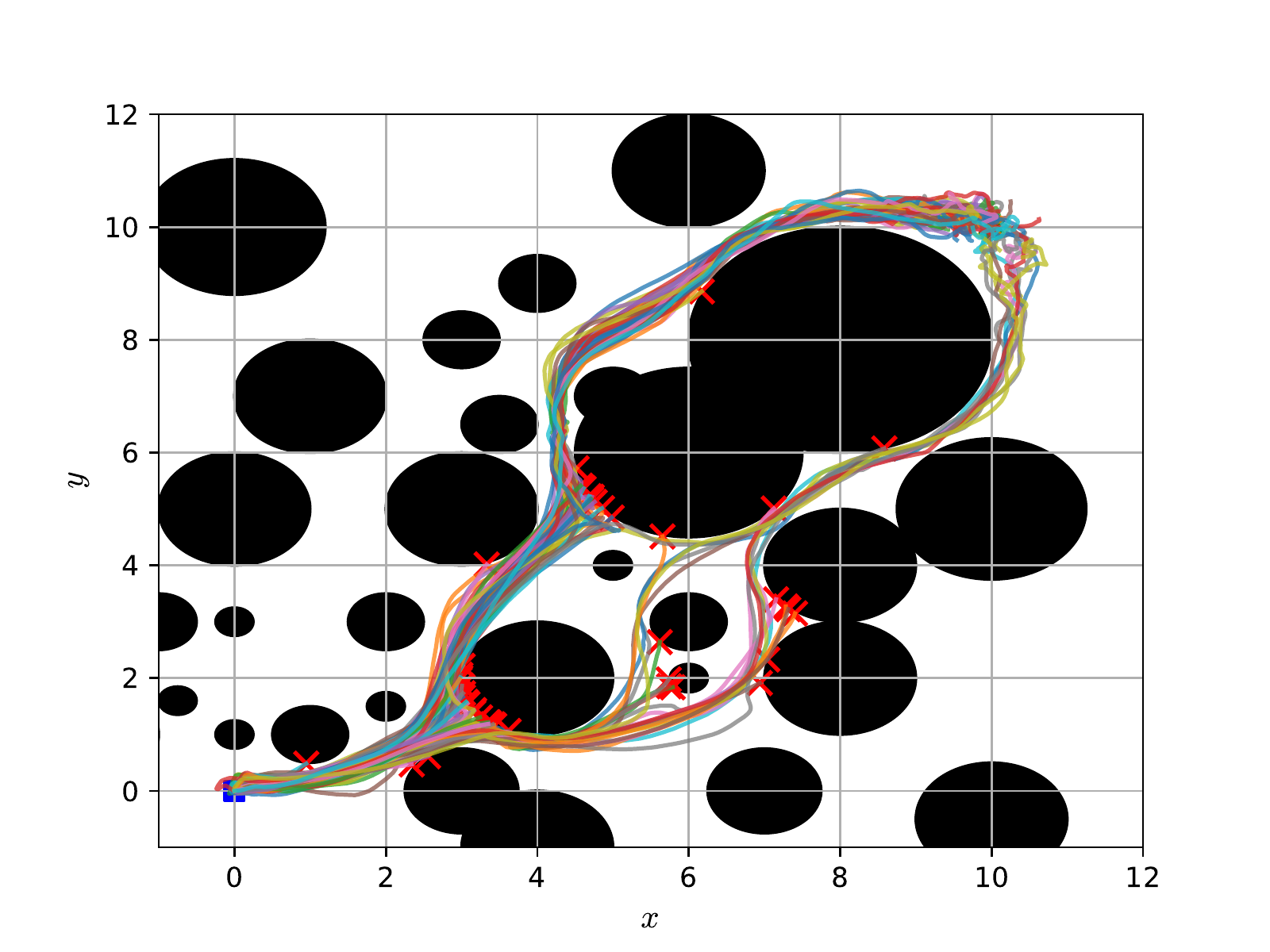}}
    \caption{Omnidirectional robot navigating through a cluttered obstacles course starting from the blue square and targeting the green filled $\boldsymbol{\times}$ under relatively high velocities disturbance using BaS-MPPI (left) and SA-RMPPI (right). Each figure shows $100$ position realizations under zero mean i.i.d. Gaussian disturbance with standard deviations of $10$ (top) and $100$ (bottom) with red $\times$ representing crashing into obstacles. As shown, SA-RMPPI realizations are more concentrated with higher safety rate, as shown in \Cref{tab: statistical comparison of point mass}, reflecting higher robustness against disturbances.}
    \label{fig: MPPI - DI in cluttered env}
\end{figure}

For this experiment, MPPI is directly optimizing the barrier state to enforce safety in MPC. We can see in \Cref{tab: statistical comparison of point mass} that the undisturbed case resulted in a near perfect success rate and low RMSE as expected. However, when the standard deviation of velocity disturbances were increased, the safety rate of the controller dropped considerably. Note, that since these are \textit{out-of-distribution} disturbances effecting the state directly, the only response the controller is able to manage is through MPC, and the ability to quickly replan. Here we see a fatal flaw in BaS-MPPI, which is that in off-nominal conditions, safety guarantees quickly dissipate. In \Cref{fig: MPPI - DI in cluttered env}, the left-most plots are each of the 100 samples of the navigation problem. We can clearly see that the narrow vertical passage in the center of the field was a common crash location, as the system did not have the ability to recover when accelerated too close to the obstacle. This issue is compounded in the higher variance example in the bottom left plot. Here, for some of the samples, the system is able to avoid the obstacles, but makes far less progress towards the goal, and ultimately crashes in the first few passageways. \Cref{fig: MPPI - DI velocity} shows the mean and standard deviation of the state velocities of the robot. BaS-MPPI experienced larger velocities and larger variance of velocity throughout the experiments, which lead to faster reaching times, but less safe control.

\subsection{Safety Augmented RMPPI (SA-RMPPI)}
\begin{figure}[h]
    %\hspace{-2.5mm}
    \subfloat{\includegraphics[trim=25 5 20 30, clip, width=.5\linewidth]{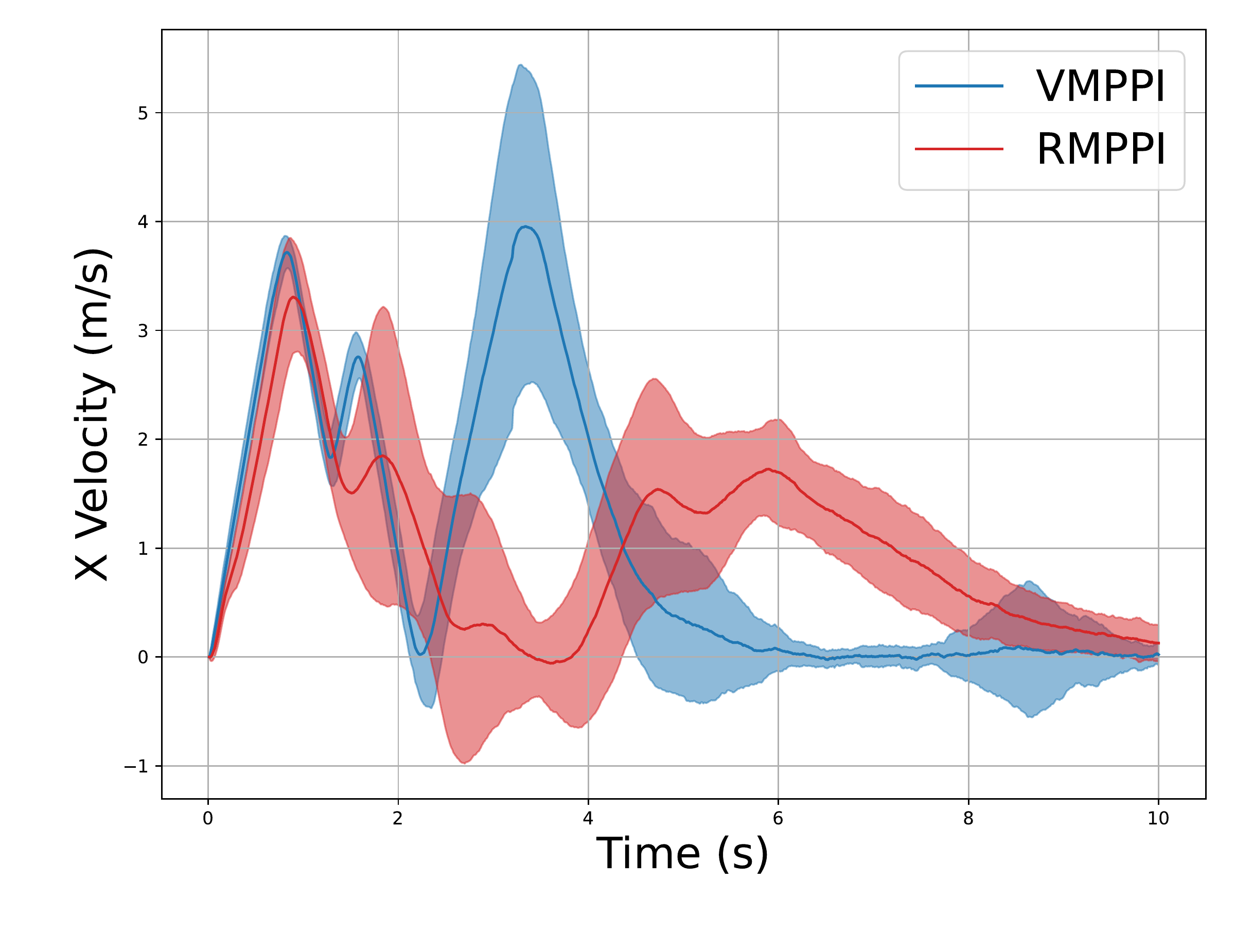}}
    %\hspace{-7mm}
    \subfloat{\includegraphics[trim=25 5 20 30, clip, width=.5\linewidth]{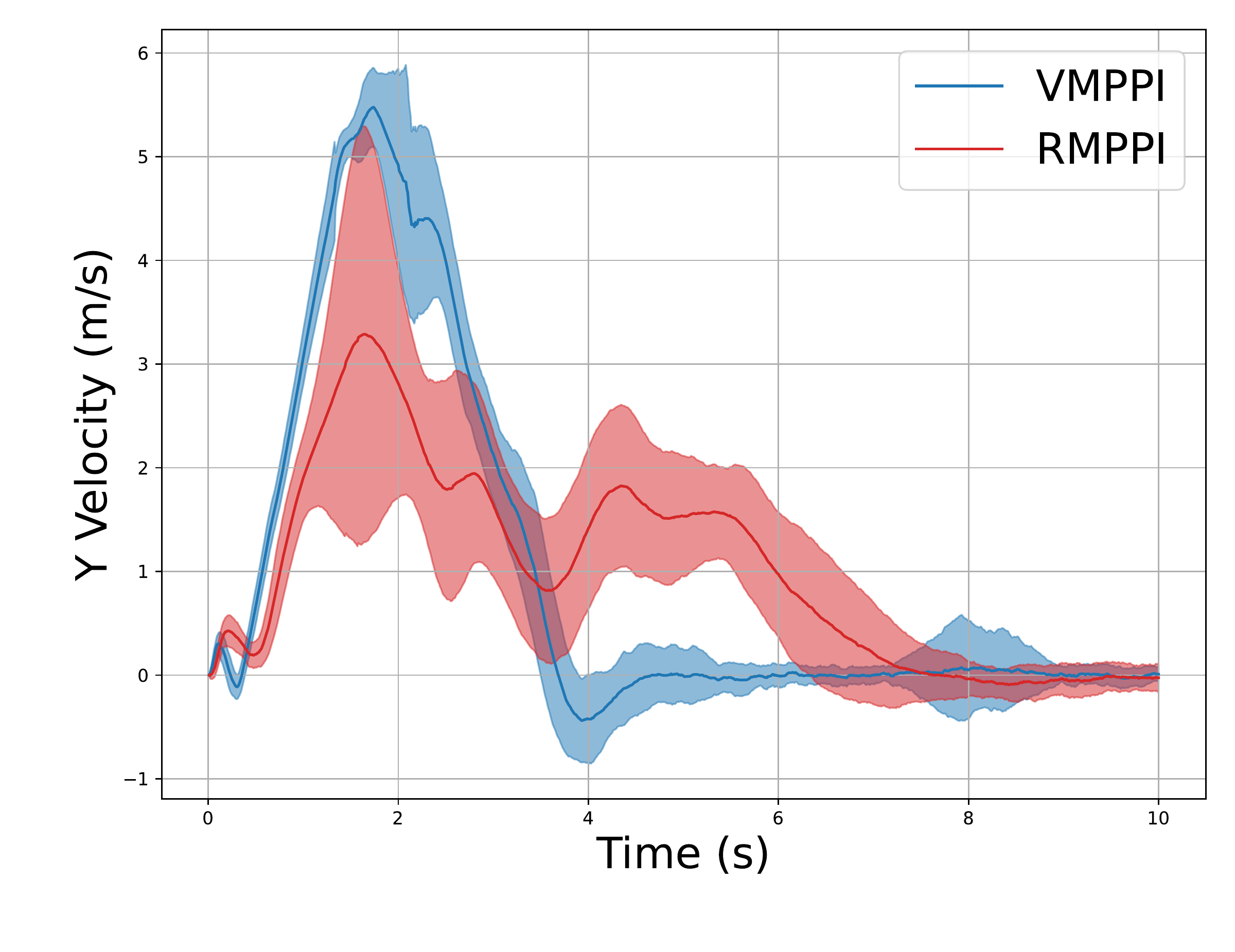}}
    \\
    %\hspace{-2.5mm}
    \subfloat{\includegraphics[trim=25 5 20 30, clip, width=.5\linewidth]{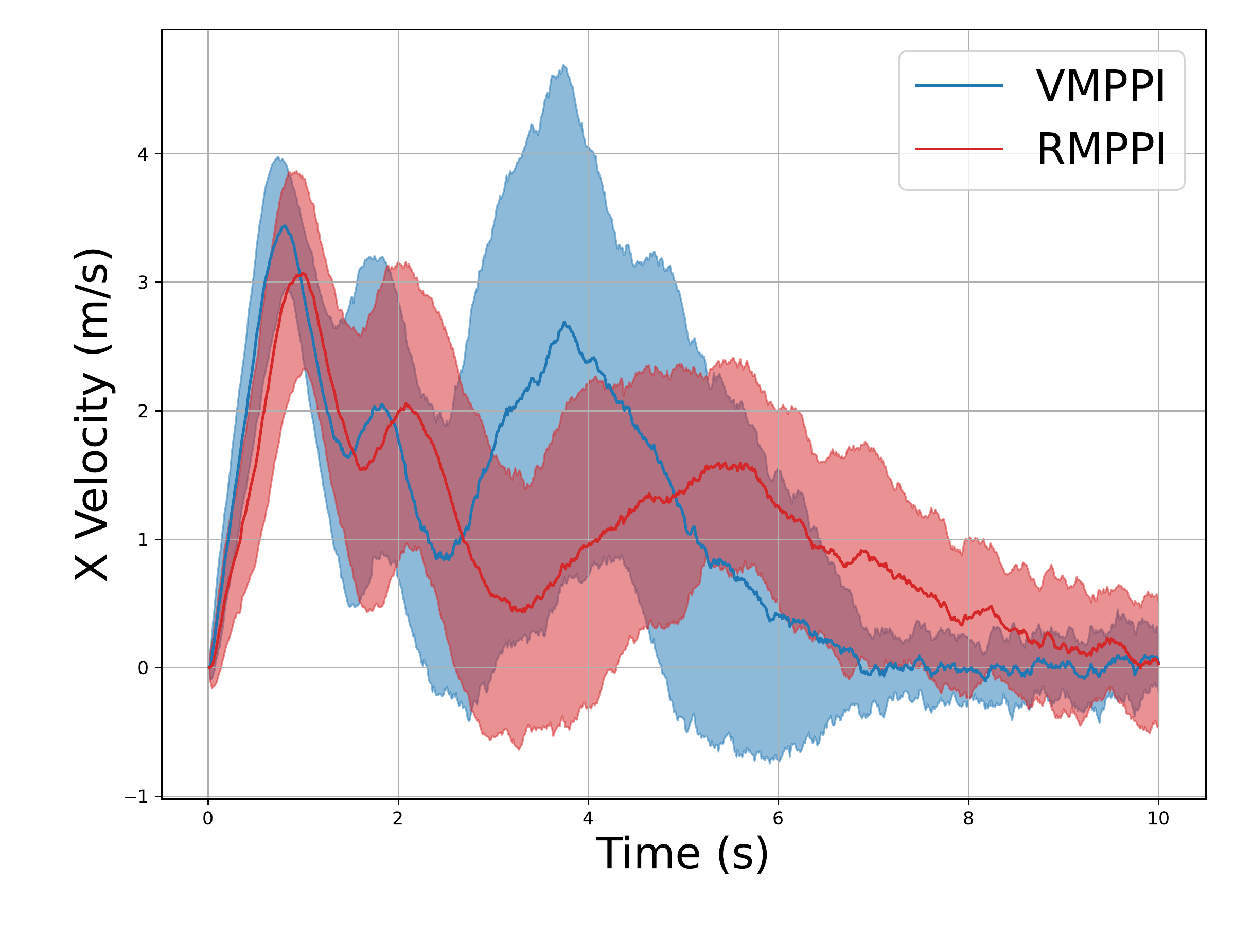}}
    %\hspace{-7mm}
    \subfloat{\includegraphics[trim=25 5 20 30, clip, width=.5\linewidth]{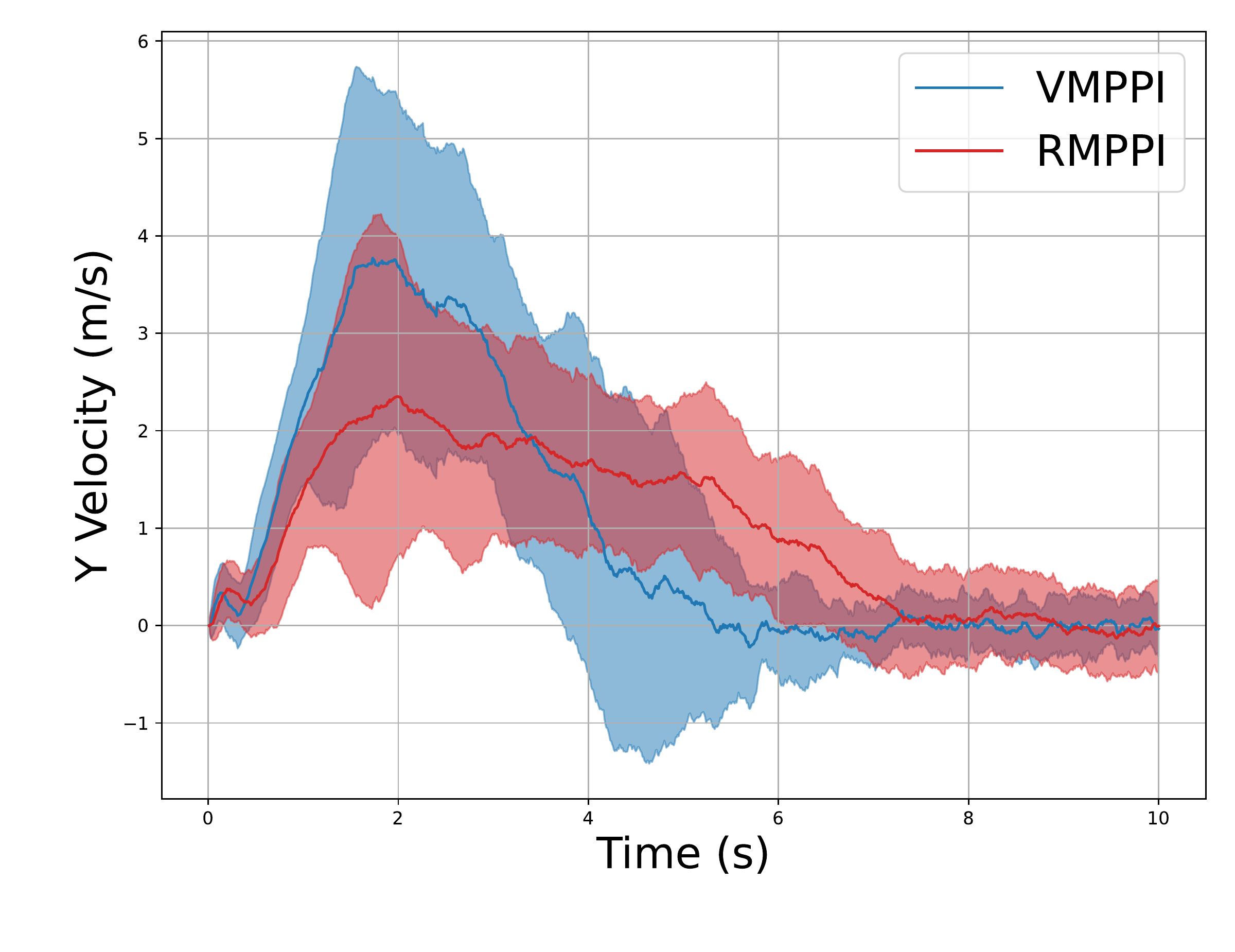}}
    \caption{Point mass robot velocities. BaS-MPPI is in blue and SA-RMPPI is in red. The top row is the mean and standard deviation error for the X (left), and Y (right) velocities for the 10 (m/s) case. The bottom row represents the 100 (m/s) case. Interestingly, the variance in velocity is higher for BaS-MPPI compared to SA-RMPPI for the bottom row, while the mean velocity is overall lower. The lower velocity and tighter variance bounds resulted in a higher safety rate. For the top row, the larger variance of SA-MPPI could be explained by the two modes seen in \Cref{fig: MPPI - DI in cluttered env} top right.}
    \label{fig: MPPI - DI velocity}
\end{figure}

With SA-RMPPI, we can see the performance gains in terms of safety in \Cref{tab: statistical comparison of point mass}. Even as the variance increases to 100, the algorithm is able to maintain a 15-24\% improvement in safety versus BaS-MPPI. This performance benefit comes a cost, one being higher computational complexity due to sampling a dual system, and two a relatively higher RMSE when approaching the goal state. This loss in performance could be due to the impact of the feedback controller in sampling, attempting to deviate samples away from obstacle even when the system is safe, or due to the \textit{conservativeness} generated by SA-RMPPI for this task. Because the system is undergoing large disturbances, we noticed in our experiments that it would a longer time horizon to meet the target, but was able to maintain safety. \Cref{fig: MPPI - DI velocity} demonstrates the lower velocity mean and tighter variance of velocity through the experiments. The velocity graph also helps explain the tighter distribution of trajectories when compared to BaS-MPPI.
\begin{figure}[h]
    \centering
    \subfloat{\includegraphics[width=.75\linewidth]{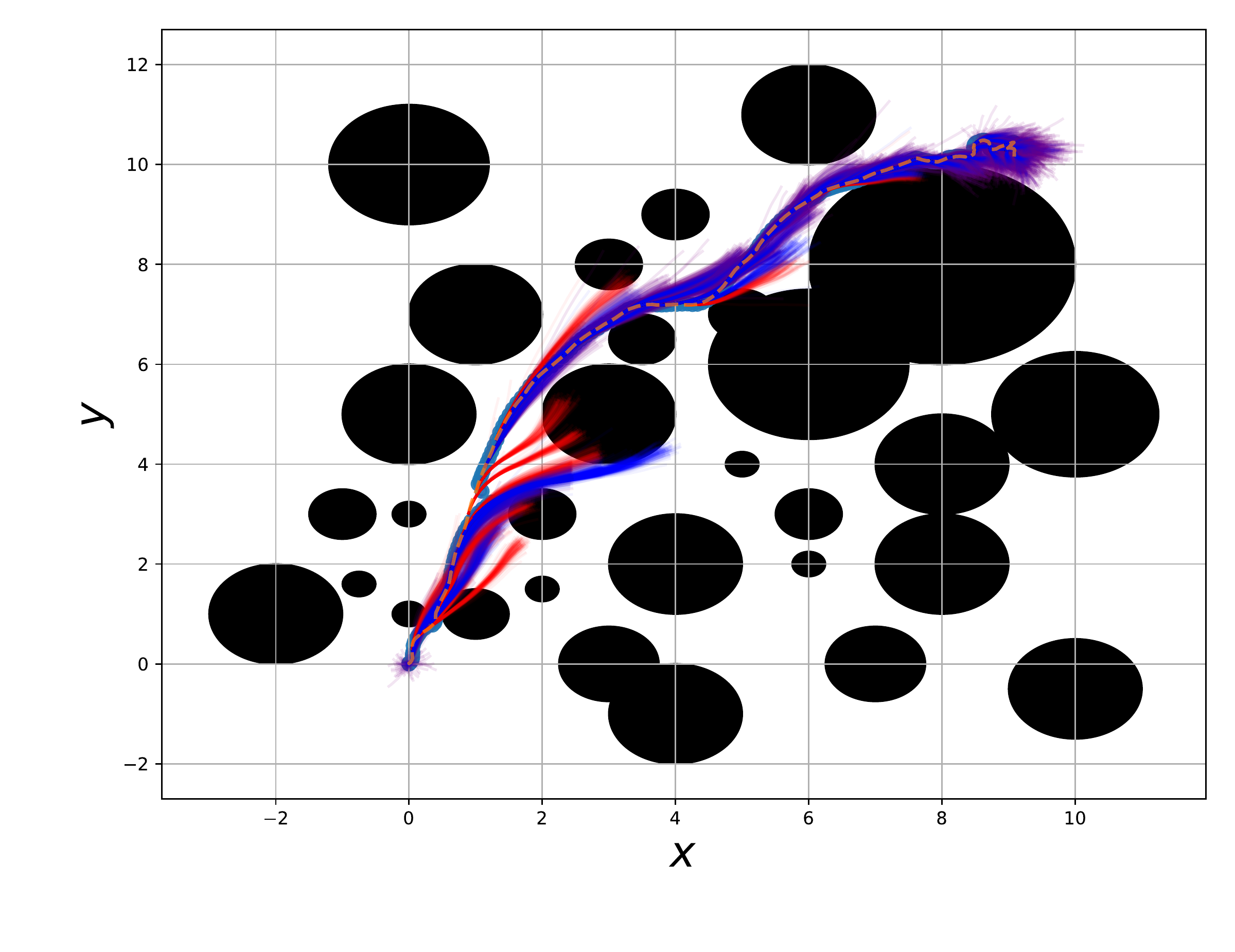}} \\
    \subfloat{\includegraphics[width=.75\linewidth]{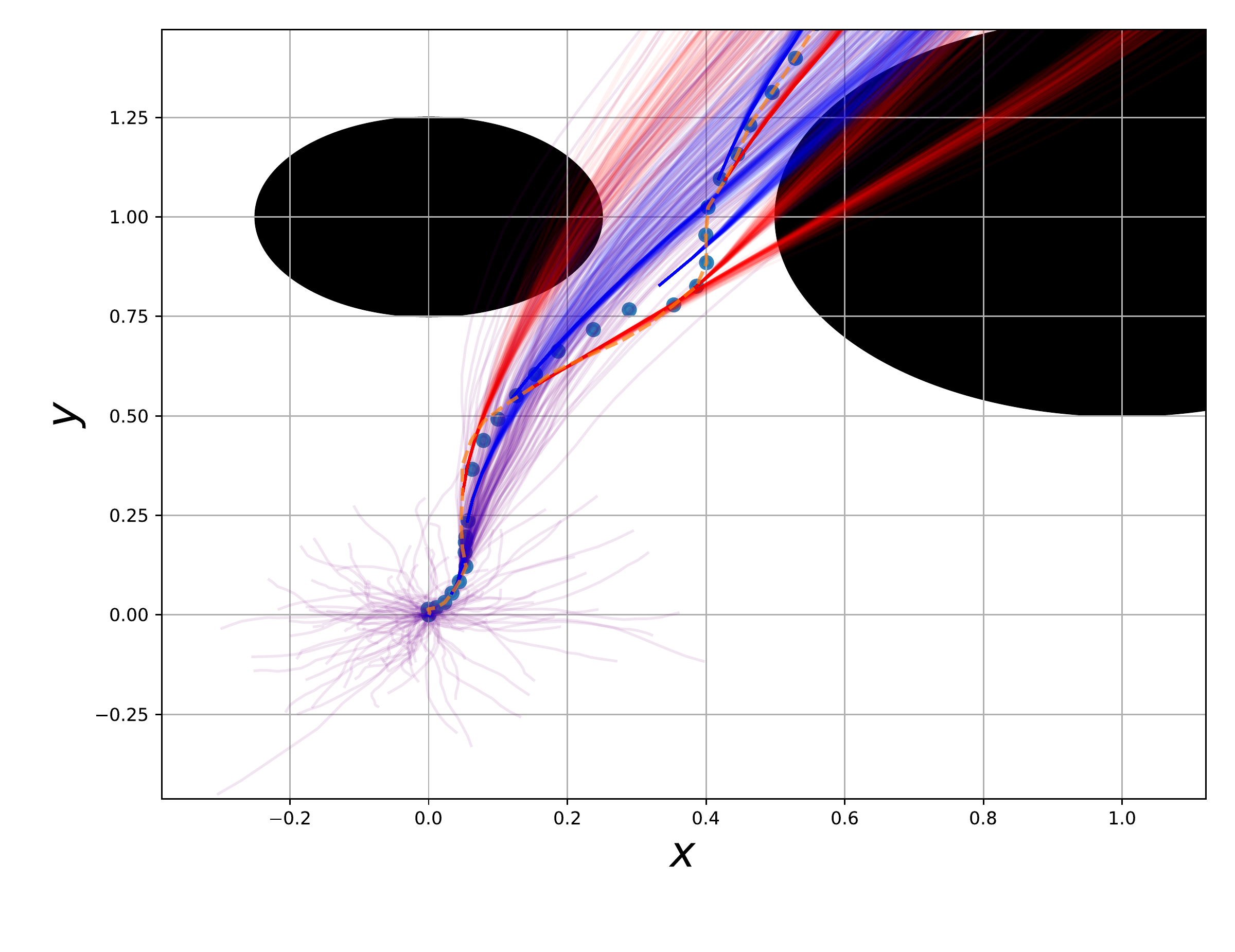}}  
    \caption{Nominal and Feedback Sampling Distributions for Out-of-distribution Noise. Blue trajectories represent the nominal samples, while red trajectories represent the samples undergoing feedback. Trajectories that appear purple imply that feedback is not activated since the trajectories are on top of one another. The blue dots represent the safe nominal state, while the. The top figures the full path of the point mass. The bottom figure shows a zoomed-in view of the initial state, with orange dashed line representing the true path. This example is with control limits of $\pm$ 50 ($\text{m}/\text{s}$) and velocity variance of 5000 ($\text{m}/\text{s}$) to illustrate spreads of trajectories and the effect of state disturbances.}
    \label{fig:RMPPI View sampling}
\end{figure}
In \Cref{fig:RMPPI View sampling}, we can see the effects of large state disturbances in SA-RMPPI. When the nominal state deviates from the real state, the feedback mechanism activates to attempt to maintain safety and return back to the nominal. In a cluttered environment, these tasks are often conflicting, and since the iLQG solution is solved about the nominal trajectory, the domain of attraction of the linear feedback gains may not be large enough to encompass all samples. We see that near the initial conditions, the feedback trajectories point away from obstacles and towards the nominal trajectory, but towards the end of the time horizon, the samples often collide with the obstacles. The safety mechanism provided by discrete barrier states relies on an accurate state linearization, and in practice this can be violated even for a simple system with highly nonlinear constraints. Note, that even though the dynamical system is a linear point mass, the barrier state dynamics are highly \textit{nonlinear}. Since the barrier state utilizes an exponential barrier function, the solution is also highly sensitive to the gradients computed along the nominal trajectory. This example shows an extreme cause of disturbances and how the system attempts to recover via safe augmented importance sampling.

\begin{table}[]
\centering
\begin{tabular}{|c|cc|cc|cc|}
\hline
\multirow{2}{*}{$\sigma^2$} & \multicolumn{2}{c|}{Safety ($\%$)}& \multicolumn{2}{c|}{RMSE}     \\ \cline{2-5} 
         & \multicolumn{1}{c|}{SA-RMPPI}       & BaS-MPPI    & \multicolumn{1}{c|}{SA-RMPPI}     & BaS-MPPI \\ \hline
0        & \multicolumn{1}{c|}{97}             &\textbf{99}  & \multicolumn{1}{c|}{0.1815} & \textbf{0.002}\\
5        & \multicolumn{1}{c|}{\textbf{91}}    &   74        & \multicolumn{1}{c|}{0.2021} & \textbf{0.018}  \\
10       & \multicolumn{1}{c|}{\textbf{87}}    &   63        & \multicolumn{1}{c|}{0.2117} & \textbf{0.0312}  \\
50       & \multicolumn{1}{c|}{\textbf{61}}    &   47        & \multicolumn{1}{c|}{0.3062} & \textbf{0.0536}  \\
100      & \multicolumn{1}{c|}{\textbf{49}}    &   31        & \multicolumn{1}{c|}{0.3634} & \textbf{0.0701}  \\
%500      & \multicolumn{1}{c|}{8}   &   \textbf{10}       & \multicolumn{1}{c|}{ 37.5}   &    90    & \multicolumn{1}{c|}{   -  }  &    85.93 \\
\hline
\end{tabular}
\caption{Comparison of SA-RMPPI and BaS-MPPI under various disturbance variance $\sigma^2$.}
\label{tab: statistical comparison of point mass}
\end{table}

\section{Conclusions}
In this work, we investigated the enforcement of safety within sampling based optimal control, namely Model Predictive Path Integral control (MPPI). We reviewed and compared three methods, namely control barrier functions (CBFs), embedded barrier states (BaS), and augmented Lagrangian optimization under moderate control disturbances to determine a suitable method for enforcing safety in sampling. As discrete barrier states showed promising performance and computational advantages over the other methods, we proposed implementing MPPI directly on the safety embedded model predictive control problem. Nonetheless, although naively optimizing with BaS provides improved performance over penalty methods with indicator functions as shown in \cite{almubarak2022safeddp}, it is subject to the same failure conditions as vanilla MPPI and is not guaranteed to be robust against system disturbances. Therefore, using barrier states in augmented importance sampling to generate safe feedback control with respect to a nominal state when sampling was proposed. Moreover, we derived computing a new bound on the free energy growth of the augmented safety embedded system, which utilizes the properties of safety critical control to localize the bound. Simulation examples under various system disturbances were presented which showed a greatly improved performance in terms of achieving the target safely.

%\begin{figure}[h]
%    \centering
%    \subfloat{\includegraphics[trim=0 20 0 0, clip, width=.75\linewidth]{Figures/RMPPI/DI_RMPPI_std100_100samples.pdf}}\\
%    \subfloat{\includegraphics[trim=0 20 0 0, clip, width=.75\linewidth]{Figures/RMPPI/DI_RMPPI_std10_100samples.pdf}}\
%    \caption{$100$ trajectories of DI with DBaS-RMPPI in a cluttered environment. Start is blue square, end is green filled $\boldsymbol{\times}$ and red $\times$ implies hitting the obstacles. Top figure is with noise in the velocities with zero mean and std of $10$ and bottom zero mean and std of $50$.}
%    \label{fig: RMPPI - DI in cluttered env}
%\end{figure}

% \section*{Acknowledgments}

%% Use plainnat to work nicely with natbib. 

\bibliographystyle{plainnat}
\bibliography{references}

\begin{thebibliography}{26}
\providecommand{\natexlab}[1]{#1}
\providecommand{\url}[1]{\texttt{#1}}
\expandafter\ifx\csname urlstyle\endcsname\relax
  \providecommand{\doi}[1]{doi: #1}\else
  \providecommand{\doi}{doi: \begingroup \urlstyle{rm}\Url}\fi

\bibitem[Agrawal and Sreenath(2017)]{agrawal2017discrete}
Ayush Agrawal and Koushil Sreenath.
\newblock \href{https://rss2017.lids.mit.edu/program/papers/22/}{Discrete
  Control Barrier Functions for Safety-Critical Control of Discrete Systems
  with Application to Bipedal Robot Navigation.}
\newblock In \emph{Robotics: Science and Systems}, volume~13. Cambridge, MA,
  USA, 2017.

\bibitem[Almubarak et~al.(2022{\natexlab{a}})Almubarak, Sadegh, and
  Theodorou]{almubarak2022safety}
Hassan Almubarak, Nader Sadegh, and Evangelos~A. Theodorou.
\newblock \href{https://arxiv.org/abs/2102.10253}{Safety Embedded Control of
  Nonlinear Systems via Barrier States}.
\newblock \emph{IEEE Control Systems Letters}, 6:\penalty0 1328--1333,
  2022{\natexlab{a}}.

\bibitem[Almubarak et~al.(2022{\natexlab{b}})Almubarak, Stachowicz, Sadegh, and
  Theodorou]{almubarak2022safeddp}
Hassan Almubarak, Kyle Stachowicz, Nader Sadegh, and Evangelos Theodorou.
\newblock \href{https://ieeexplore.ieee.org/document/9682554}{Safety Embedded
  Differential Dynamic Programming using Discrete Barrier States}.
\newblock \emph{IEEE Robotics and Automation Letters}, 2022{\natexlab{b}}.

\bibitem[Ames et~al.(2014)Ames, Grizzle, and Tabuada]{ames2014control}
Aaron~D Ames, Jessy~W Grizzle, and Paulo Tabuada.
\newblock \href{https://ieeexplore.ieee.org/document/7040372}{Control barrier
  function based quadratic programs with application to adaptive cruise
  control}.
\newblock In \emph{53rd IEEE Conference on Decision and Control}, pages
  6271--6278. IEEE, 2014.

\bibitem[Ames et~al.(2016)Ames, Xu, Grizzle, and Tabuada]{ames2016control}
Aaron~D Ames, Xiangru Xu, Jessy~W Grizzle, and Paulo Tabuada.
\newblock \href{https://ieeexplore.ieee.org/document/7782377}{Control barrier
  function based quadratic programs for safety critical systems}.
\newblock \emph{IEEE Transactions on Automatic Control}, 62\penalty0
  (8):\penalty0 3861--3876, 2016.

\bibitem[Ames et~al.(2019)Ames, Coogan, Egerstedt, Notomista, Sreenath, and
  Tabuada]{ames2019control}
Aaron~D Ames, Samuel Coogan, Magnus Egerstedt, Gennaro Notomista, Koushil
  Sreenath, and Paulo Tabuada.
\newblock \href{https://ieeexplore.ieee.org/document/8796030}{Control barrier
  functions: Theory and applications}.
\newblock In \emph{2019 18th European Control Conference (ECC)}, pages
  3420--3431. IEEE, 2019.

\bibitem[Aoyama et~al.(2021)Aoyama, Boutselis, Patel, and
  Theodorou]{Aoyama2021constddp}
Yuichiro Aoyama, George Boutselis, Akash Patel, and Evangelos~A. Theodorou.
\newblock \href{https://ieeexplore.ieee.org/document/9561530}{Constrained
  Differential Dynamic Programming Revisited}.
\newblock In \emph{2021 IEEE International Conference on Robotics and
  Automation (ICRA)}, pages 9738--9744, 2021.
\newblock \doi{10.1109/ICRA48506.2021.9561530}.

\bibitem[Berkenkamp et~al.(2017)Berkenkamp, Turchetta, Schoellig, and
  Krause]{berkenkamp2017safe}
Felix Berkenkamp, Matteo Turchetta, Angela~P Schoellig, and Andreas Krause.
\newblock
  \href{https://papers.nips.cc/paper/2017/hash/766ebcd59621e305170616ba3d3dac32-Abstract.html}{Safe
  model-based reinforcement learning with stability guarantees}.
\newblock In \emph{Proceedings of the 31st International Conference on Neural
  Information Processing Systems}, pages 908--919, 2017.

\bibitem[Gandhi et~al.(2021)Gandhi, Vlahov, Gibson, Williams, and
  Theodorou]{gandhi2021robust}
Manan~S Gandhi, Bogdan Vlahov, Jason Gibson, Grady Williams, and Evangelos~A
  Theodorou.
\newblock \href{https://ieeexplore.ieee.org/document/9349120}{Robust Model
  Predictive Path Integral Control: Analysis and Performance Guarantees}.
\newblock \emph{IEEE Robotics and Automation Letters}, 6\penalty0 (2):\penalty0
  1423--1430, 2021.

\bibitem[Garc{\i}a and Fern{\'a}ndez(2015)]{garcia2015comprehensive}
Javier Garc{\i}a and Fernando Fern{\'a}ndez.
\newblock \href{https://www.jmlr.org/papers/volume16/garcia15a/garcia15a.pdf}{A
  comprehensive survey on safe reinforcement learning}.
\newblock \emph{Journal of Machine Learning Research}, 16\penalty0
  (1):\penalty0 1437--1480, 2015.

\bibitem[Hestenes(1969)]{Hestenes1969MultiplierAG}
Magnus~R. Hestenes.
\newblock
  \href{https://link.springer.com/article/10.1007/BF00927673}{Multiplier and
  gradient methods}.
\newblock \emph{Journal of Optimization Theory and Applications}, 4:\penalty0
  303--320, 1969.

\bibitem[Koch et~al.(2019)Koch, Spies, and Bürger]{martin2019trustregions}
Martin Koch, Markus Spies, and Mathias Bürger.
\newblock \href{https://ieeexplore.ieee.org/abstract/document/8793846}{Trust
  Regions for Safe Sampling-Based Model Predictive Control}.
\newblock In \emph{2019 International Conference on Robotics and Automation
  (ICRA)}, pages 9313--9319, 2019.
\newblock \doi{10.1109/ICRA.2019.8793846}.

\bibitem[Koren and Borenstein(1991)]{koren1991potential}
Y.~Koren and J.~Borenstein.
\newblock \href{https://ieeexplore.ieee.org/document/131810}{Potential field
  methods and their inherent limitations for mobile robot navigation}.
\newblock In \emph{Proceedings. 1991 IEEE International Conference on Robotics
  and Automation}, pages 1398--1404 vol.2, 1991.
\newblock \doi{10.1109/ROBOT.1991.131810}.

\bibitem[Plancher et~al.(2017)Plancher, Manchester, and
  Kuindersma]{Plancher2017ALDDP}
Brian Plancher, Zachary Manchester, and Scott Kuindersma.
\newblock \href{https://ieeexplore.ieee.org/document/8206457}{Constrained
  unscented dynamic programming}.
\newblock In \emph{2017 IEEE/RSJ International Conference on Intelligent Robots
  and Systems (IROS)}, pages 5674--5680, 2017.
\newblock \doi{10.1109/IROS.2017.8206457}.

\bibitem[Powell(1969)]{Powell1969AMF}
M.~J.~D. Powell.
\newblock \href{https://link.springer.com/article/10.1007/BF01588967}{A method
  for nonlinear constraints in minimization problems}.
\newblock 1969.

\bibitem[Prajna(2003)]{prajna2003barrier}
Stephen Prajna.
\newblock \href{https://ieeexplore.ieee.org/document/1273063}{Barrier
  certificates for nonlinear model validation}.
\newblock In \emph{42nd IEEE International Conference on Decision and Control
  (IEEE Cat. No. 03CH37475)}, volume~3, pages 2884--2889. IEEE, 2003.

\bibitem[Prajna and Jadbabaie(2004)]{prajna2004safety}
Stephen Prajna and Ali Jadbabaie.
\newblock
  \href{https://link.springer.com/chapter/10.1007/978-3-540-24743-2_32}{Safety
  verification of hybrid systems using barrier certificates}.
\newblock In \emph{International Workshop on Hybrid Systems: Computation and
  Control}, pages 477--492. Springer, 2004.

\bibitem[Rockafellar(1974)]{Rockafellar1974AugmentedLM}
R.~Tyrrell Rockafellar.
\newblock \href{https://epubs.siam.org/doi/10.1137/0312021}{Augmented Lagrange
  Multiplier Functions and Duality in Nonconvex Programming}.
\newblock \emph{Siam Journal on Control}, 12:\penalty0 268--285, 1974.

\bibitem[Romdlony and Jayawardhana(2014)]{romdlony2014uniting}
Muhammad~Zakiyullah Romdlony and Bayu Jayawardhana.
\newblock \href{https://ieeexplore.ieee.org/document/7039737}{Uniting control
  Lyapunov and control barrier functions}.
\newblock In \emph{53rd IEEE Conference on Decision and Control}, pages
  2293--2298. IEEE, 2014.

\bibitem[Singletary et~al.(2021)Singletary, Klingebiel, Bourne, Browning,
  Tokumaru, and Ames]{singletary2020comparative}
Andrew Singletary, Karl Klingebiel, Joseph Bourne, Andrew Browning, Phil
  Tokumaru, and Aaron Ames.
\newblock \href{https://ieeexplore.ieee.org/document/9636670}{Comparative
  analysis of control barrier functions and artificial potential fields for
  obstacle avoidance}.
\newblock In \emph{2021 IEEE/RSJ International Conference on Intelligent Robots
  and Systems (IROS)}, pages 8129--8136, 2021.

\bibitem[Tao et~al.(2021)Tao, Kim, Yoon, Hovakimyan, and
  Voulgaris]{tao2021control}
Chuyuan Tao, Hunmin Kim, Hyungjin Yoon, Naira Hovakimyan, and Petros Voulgaris.
\newblock \href{https://arxiv.org/abs/2111.06974}{Control Barrier Function
  Augmentation in Sampling-based Control Algorithm for Sample Efficiency}.
\newblock \emph{arXiv preprint arXiv:2111.06974}, 2021.

\bibitem[Thananjeyan(2021)]{thananjeyan2021abc}
Brijen Thananjeyan.
\newblock
  \href{https://link.springer.com/chapter/10.1007/978-3-030-66723-8_1}{Abc-lmpc:
  Safe sample-based learning mpc for stochastic nonlinear dynamical systems
  with adjustable boundary conditions}.
\newblock In \emph{Algorithmic Foundations of Robotics XIV: Proceedings of the
  Fourteenth Workshop on the Algorithmic Foundations of Robotics 14}, pages
  1--17. Springer International Publishing, 2021.

\bibitem[Wieland and Allg{\"o}wer(2007)]{wieland2007constructive}
Peter Wieland and Frank Allg{\"o}wer.
\newblock
  \href{https://www.sciencedirect.com/science/article/pii/S1474667016355690}{Constructive
  safety using control barrier functions}.
\newblock \emph{IFAC Proceedings Volumes}, 40\penalty0 (12):\penalty0 462--467,
  2007.

\bibitem[Williams et~al.(2017)Williams, Wagener, Goldfain, Drews, Rehg, Boots,
  and Theodorou]{williams2017information}
Grady Williams, Nolan Wagener, Brian Goldfain, Paul Drews, James~M Rehg, Byron
  Boots, and Evangelos~A Theodorou.
\newblock \href{https://ieeexplore.ieee.org/document/7989202}{Information
  theoretic MPC for model-based reinforcement learning}.
\newblock In \emph{2017 IEEE International Conference on Robotics and
  Automation (ICRA)}, pages 1714--1721. IEEE, 2017.

\bibitem[Xiao and Belta(2021)]{xiao2019control}
Wei Xiao and Calin Belta.
\newblock \href{https://ieeexplore.ieee.org/abstract/document/9516971}{High
  Order Control Barrier Functions}.
\newblock \emph{IEEE Transactions on Automatic Control}, 2021.

\bibitem[Zeng et~al.(2021)Zeng, Zhang, and Sreenath]{zeng2021safety}
Jun Zeng, Bike Zhang, and Koushil Sreenath.
\newblock \href{https://ieeexplore.ieee.org/document/9483029}{Safety-critical
  model predictive control with discrete-time control barrier function}.
\newblock In \emph{2021 American Control Conference (ACC)}, pages 3882--3889.
  IEEE, 2021.

\end{thebibliography}

\end{document}